\documentclass[epj]{svjour}
\usepackage{graphics,amsmath,amssymb}
\begin{document}
\title{Quantum thermodynamic processes: A control theory for machine cycles}
\author{Jan Birjukov\inst{1}\fnmsep\thanks { e-mail: \texttt{jan.birjukow@daad-alumni.de}} \and Thomas Jahnke\inst{2} \and G\"{u}nter Mahler\inst{2}
}                     
%
%
\institute{Chair for Theoretical Physics and Applied Mathematics, Urals State Technical
             University, Mira 19, 620002 Jekaterinburg, Russia \and Institut f\"{u}r Theoretische Physik 1, Universit\"{a}t
             Stuttgart, Pfaffenwaldring 57, D-70550 Stuttgart, Germany}
\date{Received: 8 November, 2007 / Revised version: 20 May, 2008}
%

\abstract{
The minimal set of thermodynamic control parameters consists of a statistical (thermal) and a mechanical one.
These suffice to introduce all the pertinent thermodynamic variables; thermodynamic processes can then be
defined as paths on this 2-dimensional control plane. Putting aside coherence we show that for a large class
of quantum objects with discrete spectra and for the cycles considered the Carnot efficiency applies as a
universal upper bound. In the dynamic (finite time) regime renormalized thermodynamic variables allow to
include non-equilibrium phenomena in a systematic way. The machine function ceases to exist in the large
speed limit; the way, in which this limit is reached, depends on the type of cycle considered.
} 
\maketitle
\section{Introduction}

Thermodynamics~\cite{Bejan1988,TruesdellBharatha1977,TodaKuboSaito1983} has long since been supposed
to be an extremely efficient phenomenological theory, though with regard to large systems only. However,
it has recently been shown that thermodynamical properties emerge already for small quantum systems
provided they are embedded in some appropriate environment~\cite{GemmerMichelMahler2004}.
As a consequence, the idea that a single quantum object might well be described by means of
thermodynamic concepts should no longer be considered self-contradictory.
Nano-thermodynamics~\cite{Chamberlain2002,Hill2001} is becoming an emergent field that might be
relevant to many areas ranging from nano-physics to molecular biology.

Among the basic concepts of thermodynamics there are two standing out for their fundamental and practical
significance: heat and work. It is not surprising that issues concerning quantum thermodynamic
machines attract so much theoretical interest~\cite{Alicki1979,FeldmannKosloff2003,GevaKosloff1992,HenrichMahlerMichel2007,HenrichMichelMahler2006,PalaoKosloffGordon2006,QuanZhangSun2006,ScovilSchulz1959,SegalNitzan2006,TonnerMahler2005}.

Various possible levels of description can mainly be classified by the extent to which explicit
quantum effects are being taken into account. So, time dependence may either be included by
means of a classical driving system or by coupling to another full quantum mechanical subsystem~\cite{TonnerMahler2005}.
The dynamics of the machine may or may not be allowed to manifest some quantum mechanical coherence.
Anyway, in this context one should observe that the coupling to heat baths is not an unavoidable nuisance,
but an essential part of the operation of any thermodynamic machine; decoherence~\cite{HackermuellerHornberger2004}
should thus be dominant.

In this paper we intend to explore the universal limitations for the performance of a quantum
thermodynamic machine as it arises from a conceivable driving scheme and those quantum properties
of the working substance, which could survive strong decoherence: the discrete spectrum
and probabilistic character of the state. For this purpose we employ a kind of control theory
approach with two formal parameters, one related to the spectrum and another to the state of a quantum
system (cf.~\cite{FeldmannKosloff2003,GevaKosloff1992,JahnkeBirjukovMahler2007}). They are considered as the external
control. In the ideal case this is all we need to capture the essentials of thermodynamic machines,
their possible cycles and efficiencies. This simple control model is then extended to include
an internal time scale (phenomenological relaxation time) to allow for non-equilibrium effects.
This will give us a direct access to finite time thermodynamics~\cite{GevaKosloff1992,CurAhlb1975}.

Such an approach is entirely within the spirit of thermodynamics, which itself can be formulated as
a powerful control theory~\cite{TruesdellBharatha1977}. We assume that this system theoretical scheme
captures the main features of any concrete implementation, for which coherence~\cite{FeldmannKosloff2003}
does not play a major role. While there have been speculations about dramatic differences between
classical and quantum machinery~\cite{Sheehan2002}, (even claiming violations of the second law),
we come here to the opposite conclusion: Classical and quantum mechanical machines behave
essentially identical -- to an extent, which is almost unbelievable. In fact, this scale-invariance
could hardly be expected, if one took it for granted that the pertinent (thermodynamic) concepts would,
indeed, only apply in the thermodynamic limit of the system considered.

To be sure, we consider, in a sense, the ideal case. There are other important limitations in the
nano-domain: as all length scales shrink, also the possibilities of thermal isolation become severely
constrained~(cf.~\cite{Jones}). This means, e.g., that the interaction with baths of different
temperatures may no longer be assumed to be switched on and off at will. Leakage becomes
unavoidable~\cite{HenrichMahlerMichel2007,HenrichMichelMahler2006,TonnerMahler2005}.
This aspect will presently be excluded, as are any implementation issues.


\section{Model of Control}

We consider a quantum system with discrete spectrum embedded in some environment. One can imagine
three qualitatively different functions the environment could perform with respect to the system:
a mechanical control, a decohering bath and a thermal bath considered as a statistical control.

Provided the weak coupling conditions are met, we assume that the mechanical control comes into
action through a parameter-dependent spectrum of the effective Hamiltonian,
$\{E^{\mathtt{eff}}_i (\gamma)\}_{i=1}^N$, where $N$ is the number of levels. One may
regard this control as purely mechanical as long as the level occupations can survive
the corresponding spectrum disturbance. It proves to be the case in the presence of decoherence
rapid enough (time scale $\tau_{dec}$) compared with $\dot{\gamma}$~\cite{GemmerMichelMahler2004}.

This kind of decoherence can well originate from the appropriate part of the interaction between
the system and its environment. In our model we
assume such a decohering bath as a physical prerequisite of an adiabatic process. It has another
important consequence, effectively making the coarse-grained density matrix $\rho_t$ and
the momentary hamiltonian $H^{\mathtt{eff}}(\gamma_t)$ commute, excluding as well autonomous
Schr\"{o}dinger-dynamics. So we have to assume that our quantum system always stays in a mixed
state determined by the diagonal elements of $\rho$ in the respective energy representation,
$\{p_i=\rho_{ii}\}_{i=1}^N$.

In order to introduce the statistical control, we postulate the existence of a one-parameter family
of distributions, $\{\tilde{p}_i(\alpha)\}_{i=1}^N$, inherent to particular conditions of the contact
between the system and its environment. These distributions are assumed to be stable in the sense
that whenever the actual distribution $p_i$ differs from $\tilde{p}_i(\alpha)$ at some given $\alpha$,
this deviation will decay on a time scale $\tau_R \gg \tau_{dec}$ according to
\begin{equation}
\label{eq:wrelax} \dot{p}_i=-\tau_{R}^{-1}(p_i-\tilde{p}_i(\alpha)),
\end{equation}
(relaxation time approximation). We shall refer to such $\tilde{p}_i(\alpha)$ as \emph{attractor}.

Note that for $\tau_{R}^{}$ small compared with the characteristic time of enforced $\alpha$-parameter
alterations, one gets
\begin{equation}
\label{eq:wnorelax}
 p_i=\tilde{ p}_i(\alpha)
\end{equation}
at all times, i.e. the so-called quasi-static limit \cite{TruesdellBharatha1977}.

In the very general case, all $E^{\mathtt{eff}}_i (\gamma)$ had to be regarded as $N$ independent functions that
would make the model hardly tractable. In what follows we consider, instead, the special class of the spectral control
\begin{equation}
\label{eq:defspect}
E^{\mathtt{eff}}_i (\gamma)=g(\gamma)\cdot\epsilon_i,
\end{equation}
where $g$ is some monotonous function independent of $i$, and $\{\epsilon_i\}_{i=1}^N$ are to be regarded as a set
of characteristic constants. Two pertinent examples resulting in $g(\gamma)=\gamma$ and $g(\gamma)=\gamma^{-2}$
are given in Sec~\ref{subsubsec:spininfield} and Sec~\ref{subsubsec:partclinbox}. This control allows to treat
the distribution $\tilde{p}_i(\alpha)$ formally independent of $\gamma$ even if the spectrum
does explicitly enter the occupation numbers as, e.g., in the canonical case, cf.~(\ref{eq:canonic}).


\section{\label{sec:quasistat} Quasi-static limit}

The quasi-static limit plays a fundamental role in standard thermodynamics. Here it describes an
important reference scenario: the limit of perfect control over a quantum system. Both the
spectrum~(\ref{eq:defspect}) and the energy distribution~(\ref{eq:wnorelax}) are completely specified
by $\gamma$ and $\alpha$ taken to be external control parameters. Hence, as long as the thermodynamic
quantities are properly defined, one can consider an arbitrary point in the $(\alpha,\gamma)$-plane
as a thermodynamic state implemented on the system under consideration.

\subsection{Thermodynamic quantities}

As follows from the foregoing, for the system's state, corresponding in the quasi-static limit
to a certain point in the $(\alpha,\gamma)$-plane, one can define the entropy $S$ and the internal
energy $U$ \cite{TodaKuboSaito1983} :
\begin{eqnarray}
\label{eq:Sdef} S(\alpha):=-\sum \ln{\tilde{p}_i(\alpha)}\cdot \tilde{p}_i(\alpha),\\
\label{eq:Udef} U(\alpha,\gamma):=g(\gamma) \sum \epsilon_i \cdot \tilde{p}_i(\alpha);
\end{eqnarray}

From (\ref{eq:Sdef}) we observe, taking into account the
normalization condition for $\{\tilde{p}_i(\alpha)\}_{i=1}^N$, that
\begin{equation}
\label{eq:dSda}
\frac{\mathrm{d} S}{\mathrm{d} \alpha} = -\sum \ln{\tilde{p}_i}
\left(\frac{\mathrm{d} \tilde{p}_i}{\mathrm{d} \alpha}\right).
\end{equation}

In order to introduce the notion of temperature, we employ its formal
definition~\cite{TruesdellBharatha1977,TodaKuboSaito1983} as the
conjugate variable to the entropy $S$:
\begin{equation}
\label{eq:tpdef}
T:=\left(\frac{\partial U}{\partial S}\right)_{\gamma}=
   \left(\frac{\partial U}{\partial \alpha}\right)_{\gamma}
   \left(\frac{\mathrm{d} S}{\mathrm{d} \alpha}\right)^{-1}.
\end{equation}
Together with~(\ref{eq:Udef}) and~(\ref{eq:dSda}) one gets
\begin{equation}
\label{eq:tpform} T(\alpha, \gamma)
= g(\gamma) \Theta(\alpha)^{-1},
\end{equation}
where $\Theta(\alpha)$ stands for
\begin{equation}
\label{eq:thetadef} \Theta(\alpha):=-\frac{\sum \ln{\tilde{p}_i}(\mathrm{d} \tilde{p}_i/\mathrm{d} \alpha)}
{\sum \epsilon_i (\mathrm{d} \tilde{p}_i/\mathrm{d} \alpha)},
\end{equation}
a function depending solely on $\alpha$.

Consider now the total differential of the internal energy~(\ref{eq:Udef}) as a function
of these control parameters~\cite{Alicki1979}:
\begin{equation}
\label{eq:totdiffU}
\mathrm{d}U(\alpha,\gamma)=\left(\frac{\partial U}{\partial \alpha}\right)_{\gamma} \mathrm{d}\alpha+
                           \left(\frac{\partial U}{\partial \gamma}\right)_{\alpha} \mathrm{d}\gamma.
\end{equation}
Based on~(\ref{eq:tpdef}) we introduce the infinitesimal increment
of heat, $\mathrm{d}\!\!\!^{-}Q=T\mathrm{d}S$:
\begin{equation}
\label{eq:Qdef} \mathrm{d}\!\!\!^{-}Q(\alpha,\gamma):=
\left(\frac{\partial U}{\partial \alpha}\right)_{\gamma} \mathrm{d}\alpha =g(\gamma)
\sum \epsilon_i \left(\frac{\mathrm{d} \tilde{p}_i}{\mathrm{d} \alpha}\right) \mathrm{d}\alpha.
\end{equation}
As to the second term in~(\ref{eq:totdiffU}), its meaning is the internal energy increment
under constant $\alpha$, in fact under constant entropy, cf.~(\ref{eq:Sdef}). It allows us
to consider this term as work:
\begin{equation}
\label{eq:Wdef} \mathrm{d}\!\!\!^{-}W(\alpha,\gamma):=
\left(\frac{\partial U}{\partial \gamma}\right)_{\alpha}\mathrm{d}\gamma
=\frac{\mathrm{d} g}{\mathrm{d} \gamma} \sum \epsilon_i  \tilde{p}_i \, \mathrm{d}\gamma.
\end{equation}

Here, all the thermodynamic quantities have been defined based on a specific process in the control plane.
Typical measurements (of temperature $T$, say) would rather exploit the dependence of a specific observable
$A$, like the magnetization of a paramagnetic salt, on $T$.




\subsection{Specific processes}

So far as every point in the control plane $(\alpha, \gamma)$ represents some thermodynamic state,
a continuous sequence of states, a process, can be defined by a path in this plane, eventually by some
constraint on the accessible values of control parameter pairs, $f(\alpha, \gamma) = \mathtt{const}$.
For a concrete implementation the class of actually feasible processes will severely be restricted,
just like for macroscopic machines.


Within our control theory framework the simplest examples for ideal conventional processes are \emph{isentropes} and
\emph{isochores}, identified with $\alpha = \mathtt{const}$ and $\gamma = \mathtt{const}$,
respectively. It is noteworthy that along an isentrope the temperature~(\ref{eq:tpform}) is
directly proportional to $g(\gamma)$. With $ g(\gamma) = 1/\gamma$, for example, $T(\alpha, \gamma)$
will decrease with increasing $\gamma$:
\begin{equation}
T \propto \frac{1}{\gamma}
\end{equation}
This is what happens to photons in a cavity of size $\gamma = L$, a phenomenon known also from the
photon temperature reduction in our expanding universe (cosmic microwave background~\cite{Cheng2005}).
Another example is adiabatic demagnetization (magnetic cooling), with $\gamma=B$ and $g(\gamma) = \gamma$
(see Sec~\ref{subsubsec:spininfield}) resulting in $T \propto \gamma$.

Based on~(\ref{eq:tpform}), the definition of an \emph{isothermal}
process is straight forward: $T(\alpha, \gamma)=\mathtt{const}$~$(= \mathtt{T})$.
This constraint can be cast into the convenient form:
\begin{equation}
\label{eq:isoline}
g(\gamma)=\mathtt{T}\cdot\Theta(\alpha),
\end{equation}
In which $\mathtt{T}$ plays the role of a scale factor: The shape of every isotherm
is the same, depending eventually on the chosen form for $g(\gamma)$ and  $\{\tilde{p}_i(\alpha)\}_{i=1}^N$,
the attractor.

If, as a particular choice, one takes the \emph{canonical attractor}:
\begin{equation}
\label{eq:canonic} \tilde{p}_i(\alpha)=Z_{\mathcal{C}an}^{-1}
e^{-\alpha \epsilon_i}, \quad Z_{\mathcal{C}an}=\sum e^{-\alpha \epsilon_i};
\end{equation}
one finds from~(\ref{eq:thetadef}):
\begin{equation}
\label{eq:isot_can}
\Theta_{\mathcal{C}an}(\alpha)=\alpha.
\end{equation}
With $g(\gamma) = \gamma$ the isotherms~(\ref{eq:isoline}) will now be
just straight lines in the control plane, $\gamma=\alpha\mathtt{T}$ at constant
$\mathtt{T}$. With $g(\gamma) = \gamma^{-2}$ one would get $\gamma=1/\sqrt{\alpha\mathtt{T}}$,
a sort of hyperbolae (see Fig.~\ref{fig:can_isot} and Fig.~\ref{fig:can_isot2}).


\subsection{Heat and work}

Bearing in mind the forthcoming analysis of various thermodynamic cycles, which underlie heat engine or heat pump
operation, we consider the heat and work along specific processes.
Integrating~(\ref{eq:Qdef}) and~(\ref{eq:Wdef}), one gets for the isotherms $g(\gamma)=\mathtt{T}\cdot\Theta(\alpha)$:
\begin{subequations}
\label{eq:WQisotherm}
\begin{equation}
\label{subeq:Qisotherm}
Q_{\mathcal{T}}=\mathtt{T}\sum \epsilon_i \int \Theta(\alpha)\, \mathrm{d} \tilde{p}_i(\alpha);
\end{equation}
\begin{equation}
W_{\mathcal{T}}=\mathtt{T}\sum \epsilon_i \int \tilde{p}_i(\alpha) \frac{\mathrm{d}\Theta}{\mathrm{d}\alpha}\, \mathrm{d} \alpha;
\end{equation}
\end{subequations}
for the isentropes $\alpha=\mathtt{const}$:
\begin{subequations}
\label{eq:WQisentrop}
\begin{equation}
Q_{\mathcal{S}}=0;
\end{equation}
\begin{equation}
\label{subeq:Wisentrop}
W_{\mathcal{S}}=\Delta g(\gamma) \sum \epsilon_i \tilde{p}_i(\alpha);
\end{equation}
\end{subequations}
and for the isochores $\gamma=\mathtt{const}$:
\begin{subequations}
\label{eq:WQisochor}
\begin{equation}
\label{subeq:Qisochor}
Q_{\mathcal{\gamma}}=g(\gamma)\sum\epsilon_i\Delta\tilde{p}_i(\alpha);
\end{equation}
\begin{equation}
W_{\mathcal{\gamma}}=0;
\end{equation}
\end{subequations}
where $\Delta$ denotes the corresponding increment along the process line.

The sign of the heat flows $Q_{\mathcal{T}}$ and $Q_{\mathcal{\gamma}}$ is calculated
in Appendix~\ref{sec:inequ} for the case of the canonical $\tilde{p}_i(\alpha)$~(\ref{eq:canonic}). Such general sign
statements cannot be proven for the work inputs $W_{\mathcal{T;S}}$,
as their sign depends, together with the rest, on the particular choice of $\{\epsilon_i\}_{i=1}^N$.


\subsection{Carnot cycle}

Provided the RHS of~(\ref{eq:thetadef}) behaves well, it is always possible to compose a
closed path in the $(\alpha,\gamma)$-plane from two isentropes and two isotherms --
thus leading to a conventional Carnot cycle.

All the pertinent thermodynamic quantities, such as entropy~(\ref{eq:Sdef}), internal
energy~(\ref{eq:Udef}), temperature~(\ref{eq:tpform}), heat~(\ref{eq:Qdef}) and
work~(\ref{eq:Wdef}) are defined in the quasi-static limit in such a way that
Eq.~(\ref{eq:totdiffU}) turns into the Gibbsian fundamental form:
\begin{equation}
\label{subeq:gibbsian}
\mathrm{d}U= T\mathrm{d}S +\mathrm{d}\!\!\!^{-}W,
\end{equation}
irrespective of $g(\gamma)$ or the chosen kind of attractor~(\ref{eq:wnorelax}).

Thermodynamic efficiency of a heat engine cycle is defined as the ratio
$W_{\mathtt{out}}/Q_{\mathcal{T}_\mathtt{h}}$, where $W_{\mathtt{out}}=-W_{\circ}>0$ is the work output
per cycle and $Q_{\mathcal{T}_\mathtt{h}}>0$ is the heat input during the
isothermal stage at the higher of two assigned temperatures: $\mathtt{T}_\mathtt{h}>\mathtt{T}_\mathtt{c}$.
As usual, $Q_{\mathcal{T}_\mathtt{c}}<0$ is supposed to be discarded, i.e. cannot be re-used.
The validity of the Gibbsian fundamental form immediately leads to the Carnot efficiency~\cite{TruesdellBharatha1977}
\begin{equation}
\label{eq:carnot}
\eta^{\mathcal{C}}= 1-\frac{\mathtt{T}_\mathtt{c}}{\mathtt{T}_\mathtt{h}},
\end{equation}
for any heat-engine cycle of Carnot's kind in the $(\alpha,\gamma)$-plane independent
of such model details as $\{\tilde{p}_i(\alpha)\}$ or $\{\epsilon_i\}$ or $g(\gamma)$.

Thus, the Carnot efficiency as a limiting fundamental value does not even require
thermal states, only the control scheme with two parameters, as stated. The temperatures
$\mathtt{T}_\mathtt{c}, \mathtt{T}_\mathtt{h}$ would then only have a formal meaning, though.
In some sense, this is a generalization of the well-known universality established in conventional
macroscopic thermodynamics.

\begin{figure}
\includegraphics{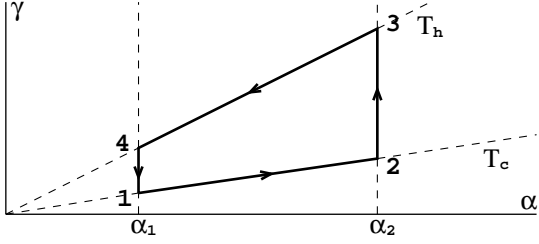}
\caption{\label{fig:can_isot}Carnot's cycle in the $(\alpha,\gamma)$-plane
associated with the canonical attractor~(\ref{eq:canonic}) and $g(\gamma)=\gamma$.
It is composed by segments of isentropes $\alpha = \alpha_{1,2}$ and isotherms
$\gamma=\alpha\mathtt{T}_{\mathtt{c},\mathtt{h}}$; $(\mathtt{T}_\mathtt{h}>\mathtt{T}_\mathtt{c})$. Arrows correspond
to heat-engine performance.}
\end{figure}

\begin{figure}
\includegraphics{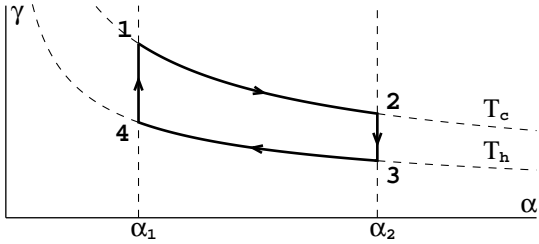}
\caption{\label{fig:can_isot2} As Fig.~\ref{fig:can_isot}, but for $g(\gamma)=\gamma^{-2}$.
The Carnot cycle is composed by segments of isentropes $\alpha = \alpha_{1,2}$ and isotherms
$\gamma=1/\sqrt{\alpha\mathtt{T}_{\mathtt{c},\mathtt{h}}}$; $(\mathtt{T}_\mathtt{h}>\mathtt{T}_\mathtt{c})$.
Arrows correspond to heat-engine performance.}
\end{figure}

In order to present an illustrative example, we explore the case of the canonical attractor~(\ref{eq:canonic}).
The corresponding Carnot cycles are sketched in Fig.~\ref{fig:can_isot} and Fig.~\ref{fig:can_isot2}.


According to~(\ref{eq:integr_ineq}), for the heat-engine performance one must drive the
cycle in Fig.~\ref{fig:can_isot} anticlockwise, while the cycle in Fig.~\ref{fig:can_isot2}
clockwise: These directions correspond to heat input for the high- and heat output for
the low-temperature isothermal stage, respectively. Treating both cycles stepwise based
on~(\ref{eq:WQisotherm}) and~(\ref{eq:WQisentrop}), one explicitly gets the Carnot
efficiency~(\ref{eq:carnot}), as expected.

Two examples for other possible choices of the attractor type are given in Appendix~\ref{sec:exot_isot}.
These examples illustrate the diversity of forms the Carnot cycles can have in the control space $(\alpha,\gamma)$
depending on the underlying attractor. In the quasi-static limit, however, all of them share the standard
rectangular form in the $(S,T)$-space (Fig.~\ref{fig:st_diag}). Also the efficiency is always
given by the Carnot value; in this sense, one cannot win anything by trying to implement exotic distribution functions.



\subsection{Otto cycle}

It is well-known that the universality established for the Carnot cycle does not carry over to other cycle
types. This means that the resulting efficiencies $\eta \le \eta^{\mathcal{C}}$ would then depend on the
actual control functions $g(\gamma)$ and $\tilde{p}_i(\alpha)$.

The Otto cycle is in some sense the most fundamental quantum thermodynamic cycle: On the isentropes, there
is only a change of the spectrum while on the isochors only the occupation numbers are changing.
Most of the theoretically discussed quantum thermodynamic machines perform, in effect, the Otto
cycle~\cite{FeldmannKosloff2003,ScovilSchulz1959,RemppMichelMahler2006}.

\begin{figure}
\includegraphics{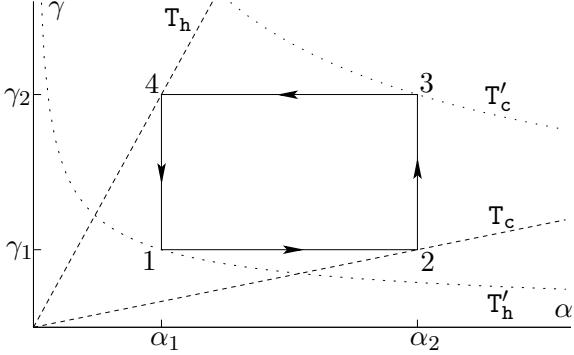}
\caption{\label{fig:otto_alpha-gamma}Otto cycle in the $(\alpha,\gamma)$-plane.
The isentropes are given by $\alpha= \mathtt{const}$ and the isochors by
$\gamma=\mathtt{const}$. The dashed lines are canonical isotherms with the highest and lowest
temperature of the cycle for $g(\gamma)=\gamma$, while the dotted ones hold for
$g(\gamma)=\gamma^{-2}$.
}
\end{figure}

Fig.~\ref{fig:otto_alpha-gamma} shows the Otto cycle in the $(\alpha,\gamma)$-plane.
As the Carnot cycle discussed above, this cycle must be driven anticlockwise or clockwise, depending
on the choice of $g(\gamma)$, in order to get the heat-engine performance.

If the function $g(\gamma)$ is increasing,~(\ref{eq:WQisentrop}) and~(\ref{eq:WQisochor}) lead to the
efficiency:
\begin{equation}
\label{eq:otto}
\eta^{\mathcal{O}}= 1-\frac{g(\gamma_1)}{g(\gamma_2)}.
\end{equation}
For a 2-level system, one gets, identifying
\begin{equation}
\Delta(\gamma)=g(\gamma)(\epsilon_2-\epsilon_1)
\end{equation}
the well-known result~\cite{Kieu2004}
\begin{equation}
\eta=1-\frac{\Delta(\gamma_1)}{\Delta(\gamma_2)},
\end{equation}
which is often mistaken as representing the Carnot efficiency~\cite{BenderBrodyMeister2002}.

With decreasing $g(\gamma)$ one must drive the cycle clockwise to get a heat-engine. The efficiency is then given by
\begin{equation}
\label{eq:otto_2}
\eta^{\prime\mathcal{O}}= 1-\frac{g(\gamma_2)}{g(\gamma_1)},
\end{equation}
In any case, the efficiency only depends on the relative compression, just like for the classical Otto cycle~\cite{Bejan1988}.

To compare these results with the efficiency of the Carnot cycle, we consider the
canonical case (\ref{eq:isoline}--\ref{eq:isot_can}) to get
\begin{subequations}
\label{eq:otto_compare}
\begin{equation}
\eta^{\mathcal{O}}= 1-\frac{\mathtt{T}_\mathtt{c}\alpha_2}{\mathtt{T}_\mathtt{h}\alpha_1}
\end{equation}
\begin{equation}
\eta^{\prime\mathcal{O}}= 1-\frac{\mathtt{T}^{\prime}_\mathtt{c}\alpha_2}{\mathtt{T}^{\prime}_\mathtt{h}\alpha_1}
\end{equation}
\end{subequations}
where $\mathtt{T}_\mathtt{c}$ ($\mathtt{T}^{\prime}_\mathtt{c}$) is the lowest temperature and $\mathtt{T}_\mathtt{h}$ ($\mathtt{T}^{\prime}_\mathtt{h}$) the highest temperature along the cycle.
As shown in Fig.~\ref{fig:otto_alpha-gamma}, $\mathtt{T}_\mathtt{c}$ is reached at point 2 and $\mathtt{T}_\mathtt{h}$ at point 4 for increasing
$g(\gamma)$, whereas for decreasing $g(\gamma)$ $\mathtt{T}^{\prime}_\mathtt{c}$ is reached at point 3 and $\mathtt{T}^{\prime}_\mathtt{h}$ at point
1 (cf.~\cite{FeldmannKosloff2003}). As one can see, the efficiency of the Otto cycle remains below the
Carnot value. Formally, the Carnot efficiency is reached for $\alpha_2=\alpha_1$, when the total work output of
the machine vanishes (cf.~\cite{ScovilSchulz1959}).

In the following we shall give two examples of spectral control:
\subsubsection{\label{subsubsec:spininfield} Example: Spin in a magnetic field}
For a spin in the magnetic field $B$, $E_i\propto B$ holds. Identifying the control parameter $\gamma$ with $B$, we have
\begin{equation}
g(\gamma)=\gamma
\end{equation}
and therefore the efficiency~(\ref{eq:otto}) simply reads
\begin{equation}
\label{eq:otto_spin}
\eta^{\mathcal{O}}_{\mathtt{spin}}= 1-\frac{\gamma_1}{\gamma_2},
\end{equation}

\subsubsection{\label{subsubsec:partclinbox} Example: Particle in a box}
Consider a particle in a box of length $L$~\cite{Sheehan2002}. The energy levels depend on the length as
$E_i\propto 1/L^2$. This length can thus be used to control the spectrum. Identifying $\gamma$ with $L$,
we have:
\begin{equation}
g(\gamma)=\frac{1}{\gamma^2}
\end{equation}
Because $g(\gamma)$ is decreasing here, we have to take~(\ref{eq:otto_2})
as the efficiency:
\begin{equation}
\label{eq:otto_box}
\eta^{\prime\mathcal{O}}_{\mathtt{box}}= 1-\left(\frac{\gamma_1}{\gamma_2}\right)^{2}
\end{equation}

\section{Driven non-equilibrium}

Now we are going to re-define the thermodynamic quantities introduced in Sec.~\ref{sec:quasistat},
allowing for deviations of the momentary distribution $\{p_i\}$ from the attractor $\{\tilde{p}_i(\alpha)\}$.
In this way we extend the scope of our consideration beyond the quasi-static limit~(\ref{eq:wnorelax})
including processes with the cycle times comparable with the relaxation time $\tau_R$
introduced in~(\ref{eq:wrelax}).

\subsection{Beyond the quasi-static limit}

According to~(\ref{eq:wrelax}), the probabilities $p_i$ no longer depend just on the momentary control $\alpha(t)$
as in the quasi-static limit; they also become dependent on their past history. Nevertheless, they are quite well
computable, as soon as $\alpha(t)$ and the initial values are defined, and one can still define the entropy and the
internal energy just by replacing $\tilde{p}_i(\alpha)$ in~(\ref{eq:Sdef}) and~(\ref{eq:Udef}) with $p_i$.

We define the non-equilibrium work $\mathrm{d}\!\!\!^{-} W^*$ as the increment of the internal energy
due to variations of the mechanical control $\gamma$:
\begin{equation}
\label{eq:non-eq-work}
\mathrm{d}\!\!\!^{-} W^*=\left( \frac{\partial U^*}{\partial \gamma}\right)_{\!\!p_i} \mathrm{d}\gamma
=\frac{\mathrm{d} g}{\mathrm{d} \gamma}\sum \epsilon_i  p_i \, \mathrm{d}\gamma;
\end{equation}
while the non-equilibrium heat $\mathrm{d}\!\!\!^{-} Q^*$ based on the first law:
\begin{equation}
\label{eq:non-eq-heat}
\mathrm{d}\!\!\!^{-} Q^*= \mathrm{d}U^*-\mathrm{d}\!\!\!^{-} W^*=g(\gamma)\sum \epsilon_i\mathrm{d}p_i.
\end{equation}
(By an asterisk we indicate the renormalization due to non-equilibrium.)

The renormalized temperature $T^*$ will be taken to remain a control variable asking for the response
of $U^*$ to a change of the statistical control $\alpha$. Note that the non-equilibrium distribution
$\{p_i\}$ can formally be represented as
\begin{equation}
\label{eq:non-eq-dist}
p_i(\alpha,\Delta p_i)=\tilde{p}_i(\alpha)+\Delta p_i,
\end{equation}
with deviations $\Delta p_i$ due to the dynamical response of the system. Being well computable
by means of the relaxation equations~(\ref{eq:wrelax}), they are, however, not under our direct control.
That is why we seek the response to a change of $\alpha$ at some given $\{\Delta p_i\}$:
\begin{equation}
\label{eq:non-eq-temp}
T^*=\left( \frac{\partial U^*}{\partial \alpha}\right)_{\gamma,\Delta p_i}
\left( \frac{\partial S^*}{\partial \alpha}\right)^{-1}_{\Delta p_i}.
\end{equation}

Such a definition results in (cf.~(\ref{eq:tpform})):
\begin{equation}
\label{eq:tpalph}
T^{*}=-g(\gamma)\frac{\sum \epsilon_i (\mathrm{d}\tilde{p}_i/\mathrm{d}\alpha)}
                     {\sum \ln{p_i}   (\mathrm{d}\tilde{p}_i/\mathrm{d}\alpha)}.
\end{equation}
The quantity $T^{*}$, determined in this way, we shall call \emph{renormalized process temperature}.

In the following we assume the time-dependence of the control parameter $\alpha(t)$ to be:
\begin{equation}
\label{eq:linalph}
\alpha=\alpha_0+\upsilon t,
\end{equation}
where $\upsilon$ may be positive or negative. Of course, there are other possible choices,
indicating that more details of the control may become important.

With the attractor $\{\tilde{p}_i(\alpha)\}$ driven
according to (\ref{eq:linalph}), the relaxation equations~(\ref{eq:wrelax}) take the form:
\begin{equation}
\label{eq:alphder} (\mathrm{d} p_i/\mathrm{d} \alpha)=
-(\upsilon\tau_{R}^{})^{-1}(p_i-\tilde{p}_i(\alpha))
\end{equation}
and yield the solutions:
\begin{equation}
\label{eq:walpha}
p_i=p_i^{(0)} e^{-\frac{\alpha-\alpha_0}{\upsilon\tau_{R}}}
   +(\upsilon\tau_{R}^{})^{-1}\int_{\alpha_0}^{\alpha}
    \tilde{p}_i(\alpha^{\prime})e^{-\frac{\alpha-\alpha^{\prime}}{\upsilon\tau_{R}}}\mathrm{d}\alpha^{\prime}.
\end{equation}
This means that, in a sense, the $p_i$ could still be considered as an explicit function of $\alpha$
as soon as their initial values $p_i^0$ and $\alpha_0$ are specified by the given process history.


\subsection{\label{subsec:renormalizedprocesstemperature} Renormalized process temperature}

In order to understand what happens with the process temperature along non-isentropic
paths in the control plane, we consider the asymptote under very slow driving,\newline
$(\upsilon\tau_{R}^{}/\Delta\alpha)\ll1$, when the following expansion
holds (see App.~\ref{sec:asymp}):
\begin{equation}
\label{eq:probexp}
p_i=\tilde{p}_i(\alpha)-\upsilon\tau_{R}^{}\frac{\mathrm{d}\tilde{p}_i}{\mathrm{d}\alpha}
+o\left[\frac{\upsilon\tau_{R}^{}}{\Delta\alpha}\right],
\end{equation}
where $\Delta\alpha$ denotes the increment of $\alpha$ in the course of the process.
Inserting this expansion into~(\ref{eq:tpalph}) and keeping terms up to first order
in $(\upsilon\tau_{R}^{}/\Delta\alpha)$, one gets:
\begin{equation}
\label{eq:tpexp}
T^{*}=T(\alpha,\gamma)\left(1+\upsilon\tau_{R}^{}\frac{\sum (1/\tilde{p}_i)(\mathrm{d}\tilde{p}_i/\mathrm{d}\alpha)^2}
                                                    {\sum \ln{\tilde{p}_i}(\mathrm{d}\tilde{p}_i/\mathrm{d}\alpha)}\right)
+o\left[\frac{\upsilon\tau_{R}^{}}{\Delta\alpha}\right],
\end{equation}
observing as well the definition~(\ref{eq:tpform}) for the temperature $T(\alpha,\gamma)$
in the quasi-static limit.

Of course, any chosen process line $f(\alpha, \gamma) = \mathtt{const}$ implies a certain
relationship between $\gamma$ and $\alpha$. In the canonical case~(\ref{eq:canonic}) and when the
process is taken to run along the line $g(\gamma)=\mathtt{T}\cdot\alpha$, which would be an isotherm
in the quasi-static limit, cf.~(\ref{eq:isoline}), Eq.~(\ref{eq:tpexp}) reduces to:
\begin{equation}
\label{eq:tpexp_can}
T^{*}(\alpha)=\mathtt{T}\left(1+\frac{\upsilon\tau_{R}^{}}{\alpha}\right)
+o\left[\frac{\upsilon\tau_{R}^{}}{\Delta\alpha}\right],
\end{equation}

For a Carnot machine cycle (Fig.~\ref{fig:can_isot} and Fig.~\ref{fig:can_isot2})
this means, compared with the quasi-static limit, that the dynamically renormalized process temperature
is increased for increasing $\alpha$ ($\upsilon > 0$, contact with a heat sink) and decreased for
decreasing $\alpha$ ($\upsilon < 0$, heat source).
\begin{figure}
\includegraphics{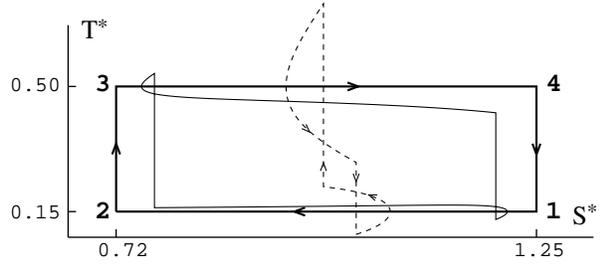}
\caption{\label{fig:st_diag} Carnot cycle in the $(S^*,T^{*})$-plane.
Both cycles from Fig.~\ref{fig:can_isot} and Fig.~\ref{fig:can_isot2}
look similar here. The rectangle denotes the perfect Carnot cycle
in the quasi-static limit $\upsilon\tau_{R} = 0$. The solid line corresponds to a stationary cyclic regime
at $\upsilon\tau_{R}^{}=0.09$; the dashed one -- at $\upsilon\tau_{R}^{}=0.9$. This is a result of numerical
simulations in accordance with $S^{*}:=-\sum \ln{p_i}\cdot p_i$ and~(\ref{eq:tpalph}), with $p_i$
under relaxation~(\ref{eq:walpha}). The set of constants $\{\epsilon_i\}_{i=1}^N$ is chosen here to be
$\{-3/2;-1/2;1/2;3/2\}$.}
\end{figure}
Temperature shifts like those following from~(\ref{eq:tpexp_can}) have \emph{ad hoc} been introduced
in Curzon and Ahlborn's finite time analysis of the heat engine efficiency~\cite{CurAhlb1975}.
Already at $\upsilon\tau_R \approx 1$, however, the $ST$ loop proper is dominated by counteracting
non-equilibrium excursions (Fig.~\ref{fig:st_diag}, cf.~\cite{GevaKosloff1992}), which indicates that
there will be no machine function left at high driving speed (cf. Sec~\ref{sec:finitetimecarnotcycle}).

One may wonder, if such a behaviour of the temperature as shown in Fig.~\ref{fig:st_diag} really can be
observed in physical systems. Indeed this seems to be the case:
Very recently such a temperature pattern was found in numerical experiments for the finite-time Carnot
cycle of a classical ideal gas by means of molecular dynamics simulations~\cite{IzumidaOkuda2008}.


\subsection{Renormalized heat and work}

Consider the renormalized heat and work along those specific processes underlying Carnot
and Otto cycles. One has to integrate now~(\ref{eq:non-eq-heat}) and~(\ref{eq:non-eq-work}) along the
corresponding lines in the $(\alpha,\gamma)$-plane. In the case of the canonical isotherm
$g(\gamma)=\alpha\mathtt{T}$ this leads to the integrals $\int^{\alpha}_{\alpha_0}\alpha'\mathrm{d}p_i$
and $\int^{\alpha}_{\alpha_0}p_i\mathrm{d}\alpha'$, with the probabilities $p_i$ under relaxation as given
by~(\ref{eq:walpha}). The latter is readily achievable directly from the relaxation
equation~(\ref{eq:alphder}):
\begin{equation}
\int^{\alpha}_{\alpha_0}p_i\,\mathrm{d}\alpha'=\int^{\alpha}_{\alpha_0}\tilde{p}_i(\alpha')\mathrm{d}\alpha'
-\upsilon\tau_{R}^{}\left(p_i-p_i^{(0)}\right),
\end{equation}
immediately followed by:
\begin{eqnarray}
\int^{\alpha}_{\alpha_0}\alpha'\,\mathrm{d}p_i&=&\int^{\alpha}_{\alpha_0}\alpha'\mathrm{d}\tilde{p}_i(\alpha')
+\alpha\left(p_i-\tilde{p}_i(\alpha)\right)\\
\nonumber
&-&\alpha_0\left(p_i^{(0)}-\tilde{p}_i(\alpha_0)\right)+\upsilon\tau_{R}^{}\left(p_i-p_i^{(0)}\right).
\end{eqnarray}

Renormalized isothermal heat and work are thus in the canonical case:
\begin{subequations}
\label{eq:WQDisotherm}
\begin{eqnarray}
\label{subeq:QDisotherm}
\nonumber Q_{\mathcal{T}}^{*}&=&Q_{\mathcal{T}}+q+q'+q'',\\
\nonumber q  &:=&\upsilon\tau_{R}^{}\mathtt{T}\sum \epsilon_i\left(p_i-p_i^{(0)}\right),\\
\nonumber q' &:=&-\alpha_0\mathtt{T}\sum \epsilon_i\left(p_i^{(0)}-\tilde{p}_i(\alpha_0)\right),\\
          q''&:=&\alpha\mathtt{T}\sum \epsilon_i\left(p_i-\tilde{p}_i(\alpha)\right);
\end{eqnarray}
\begin{eqnarray}
\label{subeq:WDisotherm}
\nonumber W_{\mathcal{T}}^{*}&=&W_{\mathcal{T}}+w,\\
          w &:=&-\upsilon\tau_{R}^{}\mathtt{T}\sum \epsilon_i\left(p_i-p_i^{(0)}\right)=-q;
\end{eqnarray}
\end{subequations}
Again, the respective renormalized quantities are denoted by an asterisk, while the heat and work
in the quasi-static limit (right hand side) are defined according to~(\ref{eq:WQisotherm}) with
$\Theta(\alpha)=\alpha$.

Integration along the isentropic line $\alpha=\mathtt{const}$ yields zero for the heat and
$\Delta g\sum\epsilon_i p_i^{(0)}$ for the work, due to the conservation (ideal in our model)
of the respective distribution~---
the results are similar to those obtained in the quasi-static limit~(\ref{eq:WQisentrop}):
\begin{subequations}
\label{eq:WQDisentrop}
\begin{eqnarray}
Q_{\mathcal{S}}^{*}&=&0;\\
\label{subeq:WDisentrop}
W_{\mathcal{S}}^{*}&=&W_{\mathcal{S}}+w',\nonumber\\
                 w'&:=&\Delta g(\gamma)\sum\epsilon_i \left(p_i^{(0)}-\tilde{p}_i(\alpha_0)\right);
\end{eqnarray}
\end{subequations}

Finally, for the isochores $\gamma=\mathtt{const}$ one gets:
\begin{subequations}
\label{eq:WQDisochor}
\begin{eqnarray}
\label{subeq:QDisochor}
\nonumber Q_{\mathcal{\gamma}}^{*}&=&Q_{\mathcal{\gamma}}+\tilde{q}+\tilde{\tilde{q}},\\
\nonumber \tilde{q}&:=&-g(\gamma)\sum \epsilon_i\left(p_i^{(0)}-\tilde{p}_i(\alpha_0)\right),\\
          \tilde{\tilde{q}}&:=&g(\gamma)\sum \epsilon_i\left(p_i-\tilde{p}_i(\alpha)\right);\\
W_{\mathcal{\gamma}}^{*}&=&0;
\end{eqnarray}
\end{subequations}
where $Q_{\mathcal{\gamma}}^{*}$, as well as $W_{\mathcal{S}}^{*}$ before, is formally expressed
through its quasi-static value.


\section{Finite-time thermodynamic cycles}

In order to realize a cyclic process in the $(\alpha,\gamma)$-control plane, we have to assign
$\alpha(t)$ for both running directions between the turning points $\alpha_1$ and $\alpha_2$.
We allow for different driving speeds, namely:
\begin{equation}
\label{eq:twospeeds}
\alpha_{1\to 2}(t)=\alpha_1+\kappa t; \qquad \alpha_{2\to 1}(t)=\alpha_2-\lambda\kappa t,
\end{equation}
with $\kappa, \lambda > 0$.

Taking~(\ref{eq:walpha}) with $\upsilon=\kappa$ and $\upsilon=-\lambda\kappa$, respectively,
one comes up with the following distributions at the turning points in the stationary cyclic regime:
\begin{subequations}
\label{eq:turningpoints}
\begin{eqnarray}
\label{subeq:turningpoint1}
p_i^{(1)}&=&\left(e^{\frac{\Delta\alpha}{\lambda\kappa\tau_{R}}}-e^{-\frac{\Delta\alpha}{\kappa\tau_{R}}}\right)^{-1}\cdot
(\kappa\tau_{R})^{-1}\\
\nonumber
&\times&\int_{\alpha_1}^{\alpha_2}
\tilde{p}_i(\alpha')\left(e^{-\frac{\alpha_2-\alpha'}{\kappa\tau_{R}}}
+\frac{1}{\lambda}\,e^{\frac{\alpha_2-\alpha'}{\lambda\kappa\tau_{R}}}\right)
\mathrm{d}\alpha';
\end{eqnarray}
\begin{eqnarray}
\label{subeq:turningpoint2}
p_i^{(2)}&=&\left(e^{\frac{\Delta\alpha}{\kappa\tau_{R}}}-e^{-\frac{\Delta\alpha}{\lambda\kappa\tau_{R}}}\right)^{-1}\cdot
(\kappa\tau_{R})^{-1}\\
\nonumber
&\times&\int_{\alpha_1}^{\alpha_2}
\tilde{p}_i(\alpha')\left(e^{\frac{\alpha'-\alpha_1}{\kappa\tau_{R}}}
+\frac{1}{\lambda}\,e^{-\frac{\alpha'-\alpha_1}{\lambda\kappa\tau_{R}}}\right)
\mathrm{d}\alpha';
\end{eqnarray}
\end{subequations}

\subsection{\label{sec:finitetimecarnotcycle} Finite-time Carnot cycle}

Consider now the Carnot cycle (Fig.~\ref{fig:can_isot} and Fig.~\ref{fig:can_isot2}) driven
 with the speed $\upsilon=\kappa$ on the low- and $\upsilon=-\lambda\kappa$
on the high-temperature isothermal stage according to~(\ref{eq:twospeeds}). The established cyclic regime
implies here the distribution~(\ref{subeq:turningpoint1}) at the points ``\textbf{1}'' and ``\textbf{4}''
and~(\ref{subeq:turningpoint2}) at the points ``\textbf{2}'' and ``\textbf{3}'' of the cycle~--- there
is no relaxation on the isentropes, as discussed before.

With~(\ref{eq:WQDisotherm}) we get the following correction terms to the quasi-static heat and
work along the isotherms $1\to 2$ and $3\to 4$; respectively:
\begin{eqnarray}
\label{eq:isot_qwcorrections}
q_{12}  &=&\kappa\tau_{R}^{}\mathtt{T}_\mathtt{c}\sum \epsilon_i\left(p_i^{(2)}-p_i^{(1)}\right)<0;\nonumber\\
q'_{12} &=&-\alpha_1\mathtt{T}_\mathtt{c}\sum \epsilon_i\left(p_i^{(1)}-\tilde{p}_i(\alpha_1)\right)>0;\nonumber\\
q''_{12}&=&\alpha_2\mathtt{T}_\mathtt{c}\sum \epsilon_i\left(p_i^{(2)}-\tilde{p}_i(\alpha_2)\right)>0;\nonumber\\
q_{34}  &=&\lambda(\mathtt{T}_\mathtt{h}/\mathtt{T}_\mathtt{c})\,q_{12}<0;\nonumber\\
q'_{34} &=&-(\mathtt{T}_\mathtt{h}/\mathtt{T}_\mathtt{c})\,q''_{12}<0;\nonumber\\
q''_{34}&=&-(\mathtt{T}_\mathtt{h}/\mathtt{T}_\mathtt{c})\,q'_{12}<0;\nonumber\\
w_{12}  &=&-q_{12}>0;\nonumber\\
w_{34}  &=&-q_{34}>0.
\end{eqnarray}

Eq.~(\ref{subeq:WDisentrop}) yields the correction to the quasi-static work along the isentropes
$2\to 3$ and $4\to 1$, respectively:
\begin{eqnarray}
\label{eq:isent_wcorrections}
w'_{23} &=& \alpha_2 \Delta \mathtt{T}\sum\epsilon_i \left(p_i^{(2)}-\tilde{p}_i(\alpha_2)\right)>0;\nonumber\\
w'_{41} &=&-\alpha_1 \Delta \mathtt{T}\sum\epsilon_i \left(p_i^{(1)}-\tilde{p}_i(\alpha_1)\right)>0;
\end{eqnarray}
Here we have observed $\Delta g(\gamma)=\alpha\Delta\mathtt{T}$, which follows from~(\ref{eq:isoline}) and~(\ref{eq:isot_can}).

\begin{figure}
\includegraphics{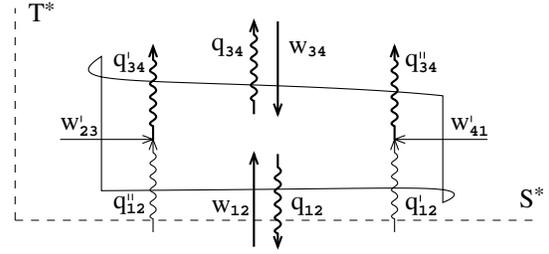}
\caption{\label{fig:ftcarnot} Corrections~(\ref{eq:isot_qwcorrections}) and~(\ref{eq:isent_wcorrections})
to the quasi-static heat and work along the stages of Carnot-like machine cycle (Fig.~\ref{fig:st_diag}).
Straight lines denote work, wavy ones -- heat. Arrows pointing into the contour indicate energy flows into
the system. (The primary, quasi-static contributions~(\ref{eq:WQisotherm}) and~(\ref{eq:WQisentrop})
providing the machine action are not shown.)}
\end{figure}

The inequalities (sign) in~(\ref{eq:isot_qwcorrections}) and~(\ref{eq:isent_wcorrections}) are proven
in Appendix~\ref{sec:inequ}. The sign of every correction term remains the same for any $\kappa$
and $\lambda$, allowing us to sketch their general scheme in Fig.~\ref{fig:ftcarnot}.
Of course, all the terms listed here relate to the corresponding cycle step as a whole,
not allowing for physical separation. Nevertheless, such a representation proves to be useful
for the purpose of the heat and work transfer analysis in various driving speed
regimes as well as in comparison with the cycles of other kinds.

In the limit of very slow driving, $(\kappa\tau_{R}^{}/\Delta\alpha) \ll 1$, one has (see App.~\ref{sec:asymp}):
\begin{eqnarray}
\label{eq:asymp_slow}
\nonumber
p_i^{(2)}-p_i^{(1)}&=&\tilde{p}_i(\alpha_2)-\tilde{p}_i(\alpha_1)+
\mathcal{O}\left[\kappa\tau_{R}^{}\right],\\
\nonumber
p_i^{(1)}-\tilde{p}_i(\alpha_1)&=&\lambda\kappa\tau_{R}^{}
\textstyle{\left(\frac{\mathrm{d}\tilde{p}_i}{\mathrm{d}\alpha}\right)_{\alpha_1}}+
o\left[\kappa\tau_{R}^{}\right],\\
p_i^{(2)}-\tilde{p}_i(\alpha_2)&=&-\kappa\tau_{R}^{}
\textstyle{\left(\frac{\mathrm{d}\tilde{p}_i}{\mathrm{d}\alpha}\right)_{\alpha_2}}+
o\left[\kappa\tau_{R}^{}\right];
\end{eqnarray}
which means that all the terms in~(\ref{eq:isot_qwcorrections}) and~(\ref{eq:isent_wcorrections})
are of first order in $\kappa\tau_{R}^{}$. In Sec.~\ref{subsec:maxpower} these results
will be used in order to reveal the fast degradation of the machine efficiency with increasing
driving speed $\kappa$.

In the limit of very fast driving, $(\Delta\alpha/\kappa\tau_{R}^{}) \ll 1$, the terms above
can be shown to be (App.~\ref{sec:asymp}):
\begin{eqnarray}
\label{eq:Corr_fast}
p_i^{(2)}-p_i^{(1)}&=&
-\frac{\Delta\alpha}{\lambda(\kappa\tau_{R}^{})^2}
\cdot\mathcal{I}_i
\textstyle
+o\left[\left(\frac{\Delta\alpha}{\kappa\tau_{R}^{}}\right)^2\right],\\
\nonumber
p_i^{(1;2)}-\tilde{p}_i(\alpha_{1;2})&=&\overline{p_i^{}}-\tilde{p}_i(\alpha_{1;2})+
\frac{1-\lambda}{\lambda\kappa\tau_{R}^{}}
\cdot\mathcal{I}_i
\textstyle
+o\left[\frac{\Delta\alpha}{\kappa\tau_{R}^{}}\right];
\end{eqnarray}
where
\begin{eqnarray}
\label{eq:I_definition}
\nonumber
\mathcal{I}_i&:=&\frac{1}{\Delta\alpha}
\int_{\alpha_1}^{\alpha_2}\!\!\left(\overline{p_i}-\tilde{p}_i(\alpha)\right)\alpha\,\mathrm{d}\alpha,\\
\overline{p_i^{}}&:=&\frac{1}{\Delta\alpha}
\int^{\alpha_2}_{\alpha_1}\!\! \tilde{ p}_i(\alpha)\mathrm{d}\alpha;
\end{eqnarray}
--~the latter being the average of $\tilde{ p}_i(\alpha)$ over the interval $[\alpha_1; \alpha_2]$.

The result~(\ref{eq:Corr_fast}) supports the intuitive expectation for the collapse of the cycle
on $(S^{*},T^{*})$-space (Fig.~\ref{fig:st_diag}) at high driving speed $\kappa$.
The only energy flows surviving $\kappa\tau_{R}^{}\to\infty$ are the total work along the
isentropic stages, $W^{*}_{23;41}$, and the quasi-static part of the work along the isotherms:
\begin{eqnarray}
\label{eq:Wnonvanish}
\nonumber
W^{*}_{23;41}=\pm\,\Delta \mathtt{T} \alpha_{2;1} \sum \epsilon_i \overline{p_i^{}};\\
W_{12;34}=\pm\,\mathtt{T}_{\mathtt{c};\mathtt{h}} \Delta\alpha \sum \epsilon_i \overline{p_i^{}}.
\end{eqnarray}

The heat flows $Q^*$ in the limit $(\Delta\alpha/\kappa\tau_{R}^{}) \ll 1$ are:
\begin{eqnarray}
\label{eq:Qfast}
\nonumber
Q^{*}_{12}&=&-\mathtt{T_\mathtt{c}} \frac{\Delta\alpha}{\kappa\tau_{R}^{}} \sum \epsilon_i \mathcal{I}_i
\textstyle
+o\left[\frac{\Delta\alpha}{\kappa\tau_{R}^{}}\right];\\
Q^{*}_{34}&=&-\mathtt{T_\mathtt{h}} \frac{\Delta\alpha}{\lambda\kappa\tau_{R}^{}} \sum \epsilon_i \mathcal{I}_i
\textstyle
+o\left[\frac{\Delta\alpha}{\kappa\tau_{R}^{}}\right],
\end{eqnarray}
negative on the both isothermal stages (App.~\ref{sec:inequ}).
Sure, there is no machine action left at $\kappa\tau_{R}^{} \gg 1$.

A comment must be made concerning the scheme in Fig.~\ref{fig:ftcarnot}. The triples $(\mathtt{q}_{12}^{\prime};
\mathtt{w}_{41}^{\prime};\mathtt{q}_{34}^{\prime\prime})$ and $(\mathtt{q}_{12}^{\prime\prime}; \mathtt{w}_{23}^{\prime};
\mathtt{q}_{34}^{\prime})$ look like effective heat-pumps, counteracting the primary machine
action along the cycle. This impression is supported by the balance relations, following from~(\ref{eq:isot_qwcorrections})
and~(\ref{eq:isent_wcorrections}):
\begin{eqnarray}
\nonumber
\mathtt{q}_{12}^{\prime}+\mathtt{w}_{41}^{\prime}+\mathtt{q}_{34}^{\prime\prime}=0;\\
\mathtt{q}_{12}^{\prime\prime}+\mathtt{w}_{23}^{\prime}+\mathtt{q}_{34}^{\prime}=0,
\end{eqnarray}
and may insinuate the idea of a crossover to a heat pump at some values of $\kappa$ and $\lambda$.
As a matter of fact, this never happens for a cycle of the kind under consideration; it is proven in
App.~\ref{sec:no_change_over} that $Q^{*}_{12}<0$ always holds~--- there is no heat absorbtion from
the cold bath as a result of the low-temperature isothermal step of the cycle.

\begin{figure}
\includegraphics{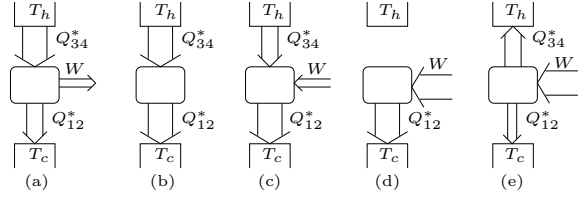}
\caption{\label{fig:carnot_wq_schematic} The development of the Carnot cycle can be divided into five steps:
For small $\kappa\tau_R$, the cycle works as a heat engine (a). At a certain velocity, the total work vanishes,
and there is only heat flowing from the hot to the cold bath (b). After that point, work has changed its sign (c).
Then there exists a velocity, where $Q_{34}^*$ vanishes. The work therefore is completely transformed into heat,
flowing into the cold bath (d). Finally, heat is flowing in the cold as well as in the hot bath (e), which
corresponds to the high driving speed limit, discussed before. Note that the size of the arrows is rescaled
at each picture: All components of heat and work tend to decrease for increasing velocity as shown in Fig.~\ref{fig:carnot_wq}.}
\end{figure}

\begin{figure}
\includegraphics{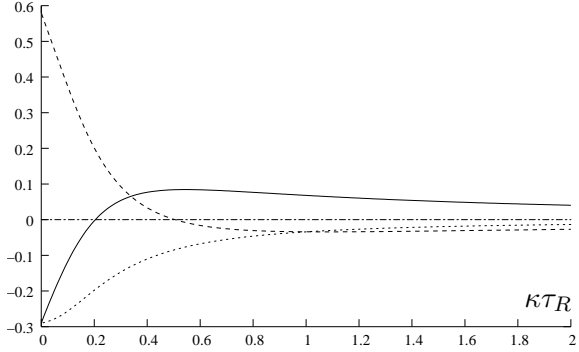}
\caption{\label{fig:carnot_wq} Behaviour of $W_\circ^*$ (solid lines), $Q_{34}^*$ (dashed lines) and $Q_{12}^*$
(dotted lines) with increasing velocity, typical for a Carnot cycle. The chosen parameters are
$\{\epsilon_i\}_{i=1}^N=\{-3/2;-1/2;1/2;3/2\}$ and $\alpha_1=0.2$, $\alpha_2=0.8$, $T_\mathtt{c}=1$,
$T_\mathtt{h}=2$. As one can see, both $W_\circ^*$ and $Q_{34}^*$ change their sign at certain finite velocities yet,
while $Q^{*}_{12}$ always stays negative.}
\end{figure}

We conclude this paragraph with graphs (Fig.~\ref{fig:carnot_wq_schematic} and Fig.~\ref{fig:carnot_wq}),
which illustrate the degradation of the Carnot cycle's machine action with the growing driving velocity.

\subsection{\label{subsec:maxpower}The efficiency at maximum power output}

In the following we restrict ourselves to corrections to the quasi-static heat and work
along the cycle stages linear in the driving speed. In the appropriate limit~(\ref{eq:asymp_slow})
eqs.~(\ref{eq:isot_qwcorrections}--\ref{eq:isent_wcorrections}) result in
\begin{subequations}
\label{eq:Corr_slow}
\begin{eqnarray}
\left\{
\begin{array}{cc}
\mathtt{w}_{12}; & \mathtt{w}_{34}
\end{array}
\right\}&=&
\left\{
\begin{array}{cc}
-\mathtt{q}_{12}; & -\mathtt{q}_{34}
\end{array}
\right\}=
\left\{
\begin{array}{cc}
\mathtt{T}_\mathtt{c}; & \lambda \mathtt{T}_\mathtt{h}
\end{array}
\right\}
\times\kappa\tau_{R} a,\nonumber\\
\textrm{where}\quad a&:=&\sum \epsilon_i \left(\tilde{p}_i(\alpha_1)-\tilde{p}_i(\alpha_2)\right)>0;\qquad
\end{eqnarray}
\begin{eqnarray}
\left\{
\begin{array}{ccc}
\mathtt{q}_{12}^{\prime}; & \mathtt{w}_{41}^{\prime}; & \mathtt{q}_{34}^{\prime\prime}\\
\end{array}
\right\}&=&
\left\{
\begin{array}{ccc}
\mathtt{T}_\mathtt{c}; & \Delta \mathtt{T}; & -\mathtt{T}_\mathtt{h}
\end{array}
\right\}
\times\lambda\kappa\tau_{R} b^{\prime},\qquad\nonumber\\
\textrm{where}\quad b^{\prime}&:= &\alpha_1\sum \epsilon_i
(-\mathrm{d}\tilde{p}_i/\mathrm{d}\alpha)_{\alpha_1}>0;\qquad
\end{eqnarray}
\begin{eqnarray}
\left\{
\begin{array}{ccc}
\mathtt{q}_{12}^{\prime\prime}; & \mathtt{w}_{23}^{\prime}; & \mathtt{q}_{34}^{\prime}\\
\end{array}
\right\}&=&
\left\{
\begin{array}{ccc}
\mathtt{T}_\mathtt{c}; & \Delta \mathtt{T}; & -\mathtt{T}_\mathtt{h}
\end{array}
\right\}
\times\kappa\tau_{R} b^{\prime\prime},\qquad\nonumber\\
\textrm{where}\quad b^{\prime\prime}&:=& \alpha_2\sum \epsilon_i
(-\mathrm{d}\tilde{ p}_i/\mathrm{d}\alpha)_{\alpha_2}>0.\qquad
\end{eqnarray}
\end{subequations}
(The inequalities are proven in Appendix~\ref{sec:inequ}.)

We assume, in addition, that the relative part of the time per cycle spent on isothermal steps remains
always the same, say $\delta$. Hence, the period $\Delta t$ for one full cycle turns out to be~\cite{CurAhlb1975}:
\begin{equation}
\label{eq:t_pro_cycle}
\Delta t=\frac{\Delta\alpha}{\delta\kappa}\frac{\lambda+1}{\lambda}.
\end{equation}

The ratio of the total work output per full cycle, $W^{*}_{\mathtt{out}}=-W^{*}_{\circ}>0$, to the heat
absorbed on the high-temperature isothermal step, $Q_{\mathcal{T}_\mathtt{h}}^{*}$, yields the efficiency $\eta^{*}$.
As follows from the scheme on Fig.~\ref{fig:ftcarnot} and~(\ref{eq:Corr_slow}), one gets:
\begin{equation}
\label{eq:not_carnot}
\eta^{*}=\frac
{W_{\mathtt{out}}-\kappa\tau_{R}^{}
\left((\mathtt{T}_\mathtt{c}+\lambda \mathtt{T}_\mathtt{h})a+\lambda\Delta \mathtt{T} b^{\prime}
+\Delta \mathtt{T} b^{\prime\prime}\right)}
{Q_{\mathcal{T}_\mathtt{h}}-\kappa\tau_{R}^{} \mathtt{T}_\mathtt{h}
\left(\lambda(a+b^{\prime})+b^{\prime\prime}\right)}
\end{equation}
One easily convinces oneself that this result is bounded from above by the quasi-static Carnot limit
\begin{equation}
\label{eq:eff_quasist}
\frac{W_{\mathtt{out}}}{Q_{\mathcal{T}_\mathtt{h}}}=\eta^{\mathcal{C}}=\frac{\Delta \mathtt{T}}{\mathtt{T}_\mathtt{h}},
\end{equation}
reached at $\kappa=0$.
The power output $\mathcal{P}^{*}=W_{\mathtt{out}}^{*}/\Delta t$ is given by
\begin{eqnarray}
\label{eq:power_out}
\mathcal{P}^{*}&=&\frac{\delta\kappa\lambda}{\Delta\alpha(\lambda+1)}\\
&\times&\Big(W_{\mathtt{out}}-\kappa\tau_{R}^{}
\big((\mathtt{T}_\mathtt{c}+\lambda \mathtt{T}_\mathtt{h})a
+\lambda\Delta \mathtt{T} b^{\prime}+\Delta \mathtt{T} b^{\prime\prime}\big)\Big)\nonumber
\end{eqnarray}
which confirms that $\mathcal{P}^{*} = 0$ for $\kappa=0$, i.e. for the maximum Carnot
efficiency $\eta^{\mathcal{C}}$.

Our aim now is to maximize $\mathcal{P}^{*}$ with respect to parameters $\kappa$ and $\lambda$ and to find
the corresponding thermodynamic efficiency~(\ref{eq:not_carnot}).

The condition for an extremum, $\partial_{\kappa} \mathcal{P}^{*}=\partial_{\lambda} \mathcal{P}^{*}=0$, yields
\begin{equation}
\label{eq:extremP}
\left\{
   \begin{array}{lcl}
      W_{\mathtt{out}}&=&2\kappa\tau_{R}^{}
           \big((\mathtt{T}_\mathtt{c}+\lambda \mathtt{T}_\mathtt{h})a+
           \lambda\Delta \mathtt{T} b^{\prime}+\Delta \mathtt{T} b^{\prime\prime}\big),\\
      W_{\mathtt{out}}&=&\kappa\tau_{R}^{}
          \big((\mathtt{T}_\mathtt{c}+\lambda \mathtt{T}_\mathtt{h})a+
          \lambda\Delta \mathtt{T} b^{\prime}+\Delta \mathtt{T} b^{\prime\prime}\big) \\
       &+&\kappa\tau_{R}^{}\lambda(\lambda+1)\big(\mathtt{T}_\mathtt{h} a + \Delta \mathtt{T} b^{\prime}\big).
   \end{array}
\right.
\end{equation}
It follows from~(\ref{eq:extremP}) that the maximum of $\mathcal{P}^{*}$ occurs at:
\begin{equation}
\label{eq:lambda}
\lambda^2=\frac{\mathtt{T}_\mathtt{c}}{\mathtt{T}_\mathtt{h}}\cdot
          \frac{1+\Delta \mathtt{T} b^{\prime\prime}/\mathtt{T}_\mathtt{c} a}
               {1+\Delta \mathtt{T} b^{\prime}/\mathtt{T}_\mathtt{h} a}.
\end{equation}

Eliminating now $Q_{\mathcal{T}_\mathtt{h}}$, $W_{\mathtt{out}}$ and $\lambda$ in~(\ref{eq:not_carnot})
by means of~(\ref{eq:eff_quasist}) and those maximum conditions, one obtains:
\begin{equation}
\label{eq:eff_extrP}
\eta_{max\mathcal{P}}^*=\eta^{\mathcal{C}}
\left(
1+\frac{1}{\sqrt{(1+\frac{\Delta \mathtt{T} b^{\prime}}{\mathtt{T}_\mathtt{h} a})
                 (1+\frac{\Delta \mathtt{T} b^{\prime\prime}}{\mathtt{T}_\mathtt{c} a})}}
\sqrt{\frac{\mathtt{T}_\mathtt{c}}{\mathtt{T}_\mathtt{h}}}
\right)^{-1}\!\!\!\!\!\!.
\end{equation}
As one can see, while the formal upper bound for $\eta_{max\mathcal{P}}^*$ is still
Carnot's $\eta^{\mathcal{C}}$, the lower bound is the celebrated Curzon-Ahlborn
efficiency~\cite{CurAhlb1975,vdBroek2005}:
\begin{equation}
\label{eq:cur_ahlb}
\eta^{\mathcal{CA}}=
1-\sqrt{\frac{\mathtt{T}_\mathtt{c}}{\mathtt{T}_\mathtt{h}}},
\end{equation}
attainable in the case, when $\Delta \mathtt{T} b^{\prime}/\mathtt{T}_\mathtt{h} a$ and
$\Delta \mathtt{T} b^{\prime\prime}/\mathtt{T}_\mathtt{c} a$ can be neglected.

Formally, it is the triples
$(\mathtt{q}_{12}^{\prime}; \mathtt{w}_{41}^{\prime}; \mathtt{q}_{34}^{\prime\prime})$ and
$(\mathtt{q}_{12}^{\prime\prime}; \mathtt{w}_{23}^{\prime}; \mathtt{q}_{34}^{\prime})$ in the
finite-time corrections to the heat and work~(\ref{eq:Corr_slow}), that makes
$\eta_{max\mathcal{P}}^*>\eta^{\mathcal{CA}}$. In fact, they originate from the
discrepancy between the steered attractor $\{\tilde{ p}_i(\alpha)\}$ and the values
of $p_i$~(\ref{eq:turningpoints}) lagging behind at the cycle turning points. It is clear, that such a
\emph{residual non-equilibrium effect} could not emerge in the purely phenomenological
thermodynamic setup by Curzon and Ahlborn~\cite{CurAhlb1975}. Nevertheless, it has been
observed for the classical ideal gas in recent numerical experiments~\cite{IzumidaOkuda2008}.

To be sure, the higher efficiency $\eta_{max\mathcal{P}}^*$ does not mean
any practical gain in the machine action here, because the corrections mentioned
above do reduce both the heat $Q_{\mathcal{T}_\mathtt{h}}$ absorbed from the
high-temperature bath and the work output $W^{*}_{\mathtt{out}}$ per cycle,
cf.~(\ref{eq:not_carnot}).

Applying Taylor's formula for $\tilde{p}_i(\alpha)$ in~(\ref{eq:Corr_slow}) provides a rough estimate
for the terms $\Delta \mathtt{T} b^{\prime}/\mathtt{T}_\mathtt{h} a$
and $\Delta \mathtt{T} b^{\prime\prime}/\mathtt{T}_\mathtt{c} a$. Observing~(\ref{eq:isot_can})
and~(\ref{eq:isoline}) as well, one gets:
\begin{equation}
\label{eq:estimations}
\frac{\Delta \mathtt{T} b^{\prime}}{\mathtt{T}_\mathtt{h} a}\approx
\frac{(\Delta g)_{\mathcal{S}_1}}{(\Delta g)_{\mathcal{T}_\mathtt{h}}};\qquad
\frac{\Delta \mathtt{T} b^{\prime\prime}}{\mathtt{T}_\mathtt{c} a}\approx
\frac{(\Delta g)_{\mathcal{S}_2}}{(\Delta g)_{\mathcal{T}_\mathtt{c}}},
\end{equation}
where $(\Delta g)_{\mathcal{T}_{\mathtt{c},\mathtt{h}}}$ and $(\Delta g)_{\mathcal{S}_{1,2}}$ stand for increments
of function $g(\gamma)$ along the corresponding isotherms and isentropes.

\begin{figure}
\includegraphics{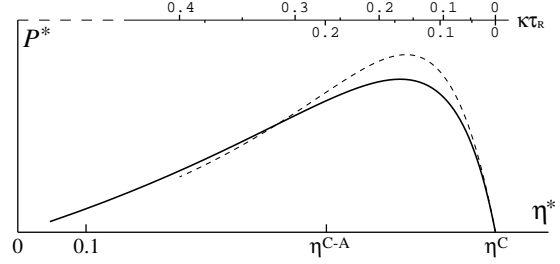}
\caption{\label{fig:efficiency} Power output
$\mathcal{P}^{*}=W_{\mathtt{out}}^{*}/\Delta t$ (arbitrary units) vs. Efficiency
$\eta^{*}=W_{\mathtt{out}}^{*} /Q_{\mathcal{T}_\mathtt{h}}^{*}$. The solid line and the
lower $\kappa\tau_{R}$-scale correspond to the linear driving speed
approximation~(\ref{eq:not_carnot}) and~(\ref{eq:power_out}), at
$\lambda=0.756$~(\ref{eq:lambda}). The dashed line and the upper
$\kappa\tau_{R}$-scale correspond to exact calculations based
on~(\ref{eq:turningpoints}), when the maximal power output appears
to be achieved at $\lambda=1.040$.}
\end{figure}

Thus, one can expect Curzon-Ahlborn's result~(\ref{eq:cur_ahlb}) for Carnot cycles
with small enough $\Delta \mathtt{T}$ and large enough $\Delta \alpha$ (eventually large enough $\Delta S$).
With the cycle parameters we have taken for our illustrations (Fig.~\ref{fig:can_isot} and Fig.~\ref{fig:can_isot2})
it is obviously not the case, however, and the optimum $\eta$, indeed, lies between the Carnot and the Curzon-Ahlborn
bounds, both for the exact and the linearized calculation (Fig.~\ref{fig:efficiency}).

\subsection{Finite-time Otto cycle}

We finally turn to the Otto machine cycle shown in Fig.~\ref{fig:otto_alpha-gamma}, driven now with finite speed.
Again we allow for two different speeds on the isochors, in accordance with~(\ref{eq:twospeeds}), and no bath coupling
and no relaxation are assumed along the isentropes. The distributions $p_i$ at the cycle's turning points ``\textbf{1}''
(``\textbf{4}'') and ``\textbf{2}'' (``\textbf{3}'') are again given by~(\ref{subeq:turningpoint1})
and~(\ref{subeq:turningpoint2}), respectively.

The heat and work along the cycle steps then follow from~(\ref{eq:WQDisentrop}) and~(\ref{eq:WQDisochor}):
\begin{eqnarray}
\label{eq:QWD_Otto}
\nonumber
Q_{12}^*&=&g(\gamma_1)\sum \epsilon_i\left(p_i^{(2)}-p_i^{(1)}\right)<0,\\
\nonumber
W_{23}^*&=&\left[g(\gamma_2)-g(\gamma_1)\right]\sum \epsilon_i {p}_i^{(2)},\\
\nonumber
Q_{34}^*&=&-g(\gamma_2)\sum \epsilon_i\left(p_i^{(2)}-p_i^{(1)}\right)>0,\\
W_{41}^*&=&-\left[g(\gamma_2)-g(\gamma_1)\right]\sum \epsilon_i {p}_i^{(1)};
\end{eqnarray}
the inequalities are proven in App.~\ref{sec:inequ}.

\begin{figure}
\includegraphics{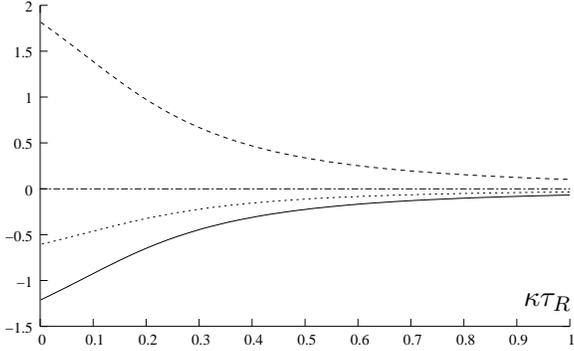}
\caption{\label{fig:otto_wq} Behaviour of $W_\circ^*$ (solid lines), $Q_{34}^*$ (dashed lines)
and $Q_{12}^*$ (dotted lines) with increasing velocity for an Otto cycle of a four level system.
The chosen parameters are $\{\epsilon_i\}_{i=1}^N=\{-3/2;-1/2;1/2;3/2\}$ and $\alpha_1=0.2$,
$\alpha_2=0.8$, $\gamma_1=1$, $\gamma_2=3$. As one can see, the ratio $-W_\circ^*/Q_{34}^*$,
which defines the efficiency, always stays the same.
}
\end{figure}

The efficiency can now easily be calculated: With increasing $g(\gamma)$ we get
\begin{equation}
\eta^{*\mathcal{O}}=\frac{-(W_{23}^*+W_{41}^*)}{Q_{34}^*}=1-\frac{g(\gamma_1)}{g(\gamma_2)}
\end{equation}
With decreasing $g(\gamma)$, one has to run the cycle clockwise to achieve heat-engine performance,
which results in the efficiency
\begin{equation}
\eta^{\prime*\mathcal{O}}=1-\frac{g(\gamma_2)}{g(\gamma_1)},
\end{equation}
i.e. exactly the same as in the quasi-static limit~(\ref{eq:otto_2}).
So, the efficiency of the Otto cycle is independent of the driving speed.
In particular, the efficiency does not depend on the power of the engine, what is
totally different from the behaviour of the Carnot cycle discussed before.
The constant efficiency does not mean, of course, that the heat and work are constant too.
All energy flows go to zero when the driving speed grows, as shown in Fig.~\ref{fig:otto_wq}.

\begin{figure}
\includegraphics{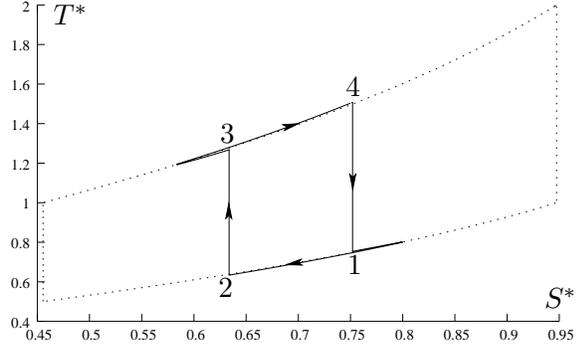}
\caption{\label{fig:otto_st}ST-diagram for an Otto cycle of a four level system with
$\{\epsilon_i\}_{i=1}^N=\{-3/2;-1/2;1/2;3/2\}$ and $\alpha_1=1$, $\alpha_2=2$, $\gamma_1=1$,
$\gamma_2=2$. The dotted lines hold for the quasi-static limit, the solid lines for
$\kappa\tau_R=0.5$. The points 1 to 4 correspond to the respective points in Fig.~\ref{fig:otto_alpha-gamma}.
}
\end{figure}
Fig.~\ref{fig:otto_st} shows a ST-diagram for an Otto cycle. In this diagram, one can also see the decrease
of the work output per cycle for increasing speed, since the encircled area decreases. But
in contrast to the Carnot cycle (Fig. \ref{fig:st_diag}), the finite speed causes spikes rather than loops.


\section{Conclusions}

We have considered a single quantum object with a discrete spectrum, an open system.
The impact of its environment has been supposed to be reducible to a parametrized
distortion of the spectrum and to a parametrized change of the occupation numbers.
The internal energy $U$ and the von Neumann entropy $S$ can then be expressed as unique functions of
these controls. The same holds for any pertinent thermodynamic quantity including those connected
not just with a state, but with a process, i.e. heat and work. While there can be no operators
underlying these quantities, they can, nevertheless, be defined for any appropriately embedded
quantum system, even down to a single spin.

Thermodynamic machines arise, if a cycle is enforced on such a two-dimensional control space.
The Carnot limit for their efficiencies has been proven and interpreted to result from
the interplay between mechanical and statistical control rather than from the spectrum details
or thermal equilibrium as such. Consequently, there is no way to violate the second law of
thermodynamics, not even by using exotic attractor states.

While coherence has been excluded here, dynamical effects have been included after defining an
internal relaxation time scale. The non-equilibrium features resulting from the finite control
speed have been incorporated in terms of renormalized thermodynamic quantities. As has been shown,
under such a non-quasi-static condition even the concept of a process temperature can still be
applicable. Compared with the quasi-static one, this renormalized temperature turns out to be lower
or higher, depending on the heat flow direction along the process.

The finite time heat and work exchange as well as the thermodynamic efficiency have been examined
in detail, analytically and numerically, for two standard machine cycles: Carnot and Otto. Though
there are some qualitative distinctions, both display rapid degradation of their machine function,
when the driving speed is increasing. Contrary to the Carnot cycle, the Otto cycle's efficiency
remains speed-independent while energy flows decrease up to zero. For the Carnot cycle the celebrated
Curzon-Ahlborn result proves to be the lower bound for the efficiency at maximum power output.

We have as yet considered a special class of time dependence for the
controls $\alpha(t)$ and $\gamma(t)$, namely, a linear one. Our approach is
meant to be some kind of a minimal model capturing the essentials of any quantum thermodynamic
machine deprived of coherence. Of course, this is not a substitute for investigations of particularized
models; nonetheless, the approach presented allows one to treat any type of cycle under arbitrary
dynamic regime on equal footing, putting aside, though, by which physical means this control might
be implemented.



\appendix

\section{\label{sec:exot_isot} Carnot cycle for non-canonical attractor}

\subsection{Nearly-uniform distribution}

As a specific attractor consider the nearly-uniform distribution
$\{\tilde{p}_i(\alpha)\}_{i=1}^N$:
\begin{equation}
\label{eq:uniform} \tilde{p}_i(\alpha)=\left\{
                    \begin{array}{rl}
                        1-\alpha, & i=1; \\
                        \alpha/(N-1), & i=2\,..\,N.
                    \end{array}
                    \right.\quad
                    \alpha \in (0,1)
\end{equation}

The temperature~(\ref{eq:tpform}) in the quasi-static
limit~(\ref{eq:wnorelax}) can be written as
\begin{equation}
T=g(\gamma) \tilde{\epsilon}
\ln^{-1}{\frac{(N-1)(1-\alpha)}{\alpha}},
\end{equation}
with
$\tilde{\epsilon}:=(N-1)^{-1}\sum^{N}_{i=2}\epsilon_i-\epsilon_1$.
From $\epsilon_1<\epsilon_{(i>1)}$ it follows that
$\tilde{\epsilon}>0$, so the area of positive temperatures on the
$(\alpha,\gamma)$-plane is restricted by $\alpha< 1-N^{-1}$.

\begin{figure}
\includegraphics{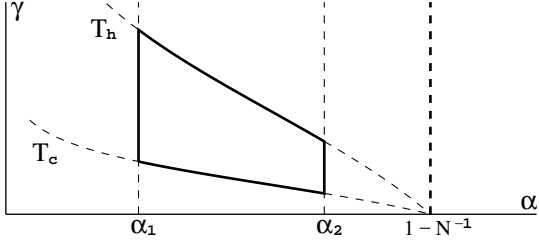}
\caption{\label{fig:uni_isot}Carnot cycle on the
$(\alpha,\gamma)$-plane associated with the nearly-uniform
distribution~(\ref{eq:uniform}) and $g(\gamma)=\gamma$. It is
composed by segments of isentropes $\alpha = \alpha_{1,2}$ and
isotherms
$\gamma=\mathtt{T}_{\mathtt{c},\mathtt{h}}\cdot\Theta_{\mathcal{U}}(\alpha)$~(\ref{eq:isot_uni});
$(\mathtt{T}_\mathtt{h}>\mathtt{T}_\mathtt{c})$.}
\end{figure}

The function $\Theta$, that defines the shape of isothermal paths
$g(\gamma)=\mathtt{T}\cdot\Theta(\alpha)$, is now
\begin{equation}
\label{eq:isot_uni}
\Theta_{\mathcal{U}}(\alpha)=\tilde{\epsilon}^{-1}\ln{\frac{(N-1)(1-\alpha)}{\alpha}};
\end{equation}
and an example of the corresponding Carnot's cycle is sketched in
Fig.~\ref{fig:uni_isot}.

\subsection{Tsallis' distribution}

The attractor $\{\tilde{ p}_i(\alpha)\}_{i=1}^N$ is now taken to be the
Tsallis' distribution~\cite{Tsallis1988}:
\begin{equation}
\label{eq:tzallis}
\begin{array}{r}
\tilde{ p}_i(\alpha)=Z_{\mathcal{T}s}^{-1}\Big(1-\alpha(q-1)\epsilon_i\Big)^{\frac{1}{q-1}},\\
Z_{\mathcal{T}s}=\sum\Big(1-\alpha(q-1)\epsilon_i\Big)^{\frac{1}{q-1}};
\end{array}
\end{equation}
with the specific parameter $q$, bounded by $q \in (0,2)$; at $q=1$ the case reduces
to the canonical one~(\ref{eq:canonic}).

Introducing the mean value $\langle\cdot\rangle$ and the covariance
\mbox{$\langle\!\langle\cdot\,;\cdot\rangle\!\rangle$} in a standard way:
\begin{equation}
\langle f\rangle:=\sum f_i \tilde{ p}_i,\quad \langle\!\langle f;g
\rangle\!\rangle:=\langle fg \rangle-\langle f\rangle\langle
g\rangle,
\end{equation}
as well as an auxiliary variable
\begin{equation}
u_i(\alpha;q)=\frac{\alpha(q-1)\epsilon_i}{1-\alpha(q-1)\epsilon_i},
\end{equation}
one can now express the isotherm shape function $\Theta_{\mathcal{T}s}(\alpha;q)$, resulting from~(\ref{eq:tpform}), as
\begin{equation}
\label{eq:isot_tza}
\Theta_{\mathcal{T}s}(\alpha;q)= \alpha \frac
{\left\langle\!\left\langle u;\ln{(1+u)}\right\rangle\!\right\rangle}
{\left\langle\!\left\langle u;  u/(1+u) \right\rangle\!\right\rangle}.
\end{equation}
Note that $\ln{(1+u)}$ and $u/(1+u)$ are both increasing functions of $u$, therefore the RHS
is always positive here.

\begin{figure}
\includegraphics{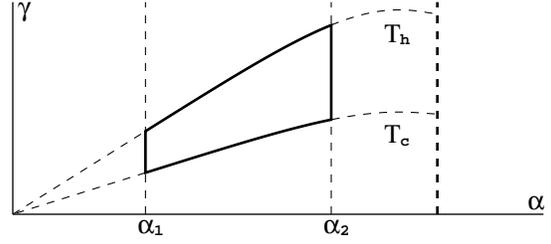}
\caption{\label{fig:tza_isot}Carnot cycle in the $(\alpha,\gamma)$-plane associated with
Tsallis' distribution~(\ref{eq:tzallis}) and $g(\gamma)=\gamma$ at $q=1.4$ and
$\{\epsilon_i\}=\{-3/2;-1/2;1/2;3/2\}$. The cycle is composed by segments of isentropes
$\alpha = \alpha_{1,2}$ and isotherms
$\gamma=\mathtt{T}_{\mathtt{c},\mathtt{h}}\cdot\Theta_{\mathcal{T}s}(\alpha)$~(\ref{eq:isot_tza});
$(\mathtt{T}_\mathtt{h}>\mathtt{T}_\mathtt{c})$. Note that the domain of definition for~(\ref{eq:tzallis}) is restricted by
\mbox{$\alpha<[\max_i{(q-1)\epsilon_i}]^{-1}$}.}
\end{figure}
An example of a Carnot cycle with isotherms of such a kind is sketched in
Fig.~\ref{fig:tza_isot}. One can see that these isotherms are very similar to the canonical
ones (cf. Fig.~\ref{fig:can_isot}) as long as $\alpha$ is small enough.
This is easily shown taking $\alpha \epsilon_i\ll 1$ for~(\ref{eq:isot_tza}).

\section{\label{sec:inequ} Inequalities}

Here we prove some inequalities concerning the heat and work flows in case of the canonical
attractor $\{\tilde{p}_i(\alpha)\}_{i=1}^N$~(\ref{eq:canonic}). Considering the first
derivative of $\tilde{p}_i(\alpha)$, one easily gets
\begin{equation}
\frac{\mathrm{d} \tilde{ p}_i(\alpha)}{\mathrm{d} \alpha}= \sum_{k}
(\epsilon_k-\epsilon_i)\tilde{ p}_k(\alpha)\tilde{ p}_i(\alpha)
\end{equation}
and, consequently:
\begin{eqnarray}
\sum_{i=1}^{N}\epsilon_i \frac{\mathrm{d}\tilde{ p}_i}{\mathrm{d}
\alpha}&=&
Z_{\mathcal{C}an}^{-2} \sum_{i,k=1}^{N} \epsilon_i (\epsilon_k-\epsilon_i) e^{-\alpha (\epsilon_k+\epsilon_i)}\nonumber\\
&=&
-Z_{\mathcal{C}an}^{-2} \sum_{i<k}(\epsilon_k-\epsilon_i)^2 e^{-\alpha (\epsilon_k+\epsilon_i)}<0.\nonumber
\end{eqnarray}
So, one gets
\begin{equation}
\label{eq:neg_deriv}
\sum \epsilon_i (\mathrm{d}\tilde{p}_i/\mathrm{d}\alpha)<0,
\end{equation}
that implies for $\alpha_1 < \alpha_2$ the following inequalities:
\begin{eqnarray}
\label{eq:differ_ineq} \sum \epsilon_i \left(\tilde{ p}_i(\alpha_1)-\tilde{p}_i(\alpha_2)\right)>0;\\
\label{eq:integr_ineq} \sum \epsilon_i \int_{\alpha_1}^{\alpha_2} \alpha\, \mathrm{d} \tilde{p}_i(\alpha)<0.
\end{eqnarray}\\

Consider the terms $\sum \epsilon_i\left(p_i^{(1;2)}-\tilde{p}_i(\alpha_{1;2})\right)$
from~(\ref{eq:isot_qwcorrections}), where the distributions $p_i^{(1;2)}$ at the cycle turning
points are defined in~(\ref{eq:turningpoints}). Integration by parts leads to:
\begin{equation}
\nonumber
\sum \epsilon_i\left(p_i^{(1;2)}-\tilde{p}_i(\alpha_{1;2})\right)=
\pm \int^{\alpha_2}_{\alpha_1}\!\!
\mathcal{F}_{1;2}\cdot
\textstyle
\left(\sum \epsilon_i \frac{\mathrm{d}\tilde{p}_i}{\mathrm{d}\alpha}\right)
\cdot \mathrm{d} \alpha,
\end{equation}
where
\begin{equation}
\nonumber
\mathcal{F}_{1;2}=
\frac{\exp{\{(\alpha-\alpha_{2;1})/\kappa\tau_{R}^{}\}}-\exp{\{(\alpha_{2;1}-\alpha)/\lambda\kappa\tau_{R}^{}\}}}
{\,\exp{\{\mp\Delta\alpha/\kappa\tau_{R}^{}\}}-\exp{\{\pm\Delta\alpha/\lambda\kappa\tau_{R}^{}\}}}.
\end{equation}
Observing $\mathcal{F}_{1;2}>0$ and~(\ref{eq:neg_deriv}), one comes up with:
\begin{equation}
\label{eq:initial_phi}
\sum \epsilon_i\left(p_i^{(1;2)}-\tilde{p}_i(\alpha_{1;2})\right)\lessgtr 0;
\end{equation}

For the next inequality we need the following\\

\noindent
\emph{\underline{Lemma}: For any two monotone decreasing functions $f(x)$ and $h(x)$
the inequality $\int_a^b h(x)f(x) \mathrm{d}x>0$ follows from $\int_a^b h(x) \mathrm{d}x=0$.}\\

\noindent
\emph{\underline{Proof}: Define $H(x)=\int_a^x h(x^{\prime}) \mathrm{d}x^{\prime}$ with the obvious
property: $H(x)>H(a)=H(b)=0$ for $a<x<b$. Applying integration by parts, one gets:
\begin{eqnarray}
\int_a^b\!\!h(x)f(x)\mathrm{d}x&=&\left[H(x)f(x)\right]_a^b-\int_a^b\frac{\mathrm{d}f}{\mathrm{d}x}H(x)\mathrm{d}x\nonumber\\
&=&-\int_a^b \frac{\mathrm{d}f}{\mathrm{d}x}H(x) \mathrm{d}x>0,\nonumber
\end{eqnarray}
observing $(\mathrm{d}f/\mathrm{d}x)<0$ and $H(x)>0$.}\\

Consider now the term $\sum \epsilon_i\left(p_i^{(2)}-p_i^{(1)}\right)$ which occurs in RHS
of~(\ref{eq:isot_qwcorrections}) and~(\ref{eq:QWD_Otto}). With~(\ref{eq:turningpoints}) it reduces to
\begin{eqnarray}
\label{eq:calculation_inequ_otto}
\sum\epsilon_i\left(p_i^{(2)}-p_i^{(1)}\right)&=&\frac{1}{\kappa\tau_R^{}}
\left\{1-e^{\frac{\Delta\alpha}{\kappa\tau_R^{}}\frac{\lambda+1}{\lambda}}\right\}^{-1}\\
    &\times& \int_{\alpha_1}^{\alpha_2}
    h(\alpha)\left(\sum\epsilon_i\tilde{p}_i(\alpha)\right)\mathrm{d}\alpha;\nonumber
\end{eqnarray}
with
\begin{equation}
\nonumber
h(\alpha)=e^{\frac{\alpha-\alpha_1}{\kappa\tau_R^{}}}
\left(1-e^{\frac{\Delta\alpha}{\lambda\kappa\tau_R^{}}}\right)+
\frac{1}{\lambda}e^{\frac{\alpha_2-\alpha}{\lambda\kappa\tau_R^{}}}
\left(e^{\frac{\Delta\alpha}{\kappa\tau_R^{}}}-1\right).
\end{equation}

One can easily check that both $\int_{\alpha_1}^{\alpha_2}h(\alpha)\mathrm{d}\alpha=0$
and $(\mathrm{d}h/\mathrm{d}\alpha)<0$ hold. It follows from~(\ref{eq:neg_deriv}) that
$\sum\epsilon_i\tilde{p}_i(\alpha)$ is a decreasing function of $\alpha$ as well.
The lemma proven above ensures the integral in~(\ref{eq:calculation_inequ_otto}) to be positive and
consequently:
\begin{equation}
\label{eq:otto_heat}
    \sum \epsilon_i\left(p_i^{(2)}-p_i^{(1)}\right)<0.
\end{equation}

Consider, at last, the term $\sum \epsilon_i \mathcal{I}_i$ from~(\ref{eq:Qfast}). The integral
$\mathcal{I}_i$ is defined in~(\ref{eq:I_definition}) and reduces to:
\begin{equation}
\int^{\alpha_2}_{\alpha_1}\!\!\! \left(\overline{ p_i^{}}-\tilde{
p}_i(\alpha)\right)\alpha\mathrm{d}\alpha=
\big(\!\int^{\overline{\alpha}}_{\alpha_1}\!\!\!+\int^{\alpha_2}_{\overline{\alpha}}\!\!\!\big)
(\overline{\alpha}-\alpha)\tilde{
p}_i(\alpha)\mathrm{d}\alpha, \nonumber
\end{equation}
where $\overline{\alpha}=(\alpha_2+\alpha_1)/2$. It follows from the mean-value theorem that
\begin{eqnarray}
\big(\!\int^{\overline{\alpha}}_{\alpha_1}\!\!\!+\int^{\alpha_2}_{\overline{\alpha}}\!\!\!\big)
(\overline{\alpha}-\alpha)\left(\sum \epsilon_i\tilde{
p}_i(\alpha)\right)\mathrm{d}\alpha&&
\nonumber\\
=\sum \epsilon_i\tilde{ p}_i(\theta_1)
\int^{\overline{\alpha}}_{\alpha_1}\!\!
(\overline{\alpha}&-&\alpha)\mathrm{d}\alpha\nonumber\\
+\sum \epsilon_i\tilde{ p}_i(\theta_2)
\int^{\alpha_2}_{\overline{\alpha}}\!\!
(\overline{\alpha}&-&\alpha)\mathrm{d}\alpha\nonumber\\
=\textstyle{\sum \epsilon_i \left(\tilde{ p}_i(\theta_1)-\tilde{
p}_i(\theta_2)\right)
\int_0^{\overline{\alpha}}u\,\mathrm{d}u},&&\nonumber
\end{eqnarray}\\
where $\alpha_1<\theta_1<\overline{\alpha}<\theta_2<\alpha_2$. With~(\ref{eq:differ_ineq})
one gets:
\begin{equation}
\Delta\alpha \sum \epsilon_i \mathcal{I}_i=\sum \epsilon_i\int^{\alpha_2}_{\alpha_1}\!\!
\left(\overline{p_i^{}}-\tilde{p}_i(\alpha)\right)\alpha\,\mathrm{d}\alpha>0.
\end{equation}

\section{\label{sec:asymp} Asymptotes}

To get the asymptotic solution of the relaxation equation~(\ref{eq:alphder}) in the slow driving
limit $(\upsilon\tau_{R}^{}/\Delta\alpha) \ll 1$, we start an iterative procedure: For this purpose
we rewrite~(\ref{eq:alphder}) as
\begin{equation}
\label{eq:DE_for_iter}
p_i=\tilde{p}_i(\alpha)-\upsilon\tau_R \left(\mathrm{d} p_i/\mathrm{d} \alpha\right)\,;
\end{equation}

The zeroth approximation is given by $p_i\approx\tilde{p}_i(\alpha)$. The first order
is obtained by the corresponding substitution on the right hand side of~(\ref{eq:DE_for_iter}):
\begin{equation}
p_i\approx\tilde{p}_i-\upsilon\tau_R \left(\mathrm{d} \tilde{p}_i/\mathrm{d} \alpha\right)\,.
\end{equation}

This result, in turn, will be used, in the next iteration step:
\begin{equation}
\label{eq:p_approx}
p_i\approx\tilde{p}_i-\upsilon\tau_R \left(\mathrm{d} \tilde{p}_i/\mathrm{d} \alpha\right)
+(\upsilon\tau_R)^2 \left(\mathrm{d}^2 \tilde{p}_i/\mathrm{d} \alpha^2\right)\,,
\end{equation}
and so on.\\

Consider now the asymptotes for the distributions $p_i^{(1;2)}$ at the cycle's turning points~(\ref{eq:turningpoints}),
first, in the limit of very slow driving, here $(\kappa\tau_{R}^{}/\Delta\alpha) \ll 1$. Both
$p_i^{(1)}$ and $p_i^{(2)}$ can be regarded as particular solutions of Eq.~(\ref{eq:DE_for_iter})
obtained at the specially chosen parameters and initial values. So, $p_i^{(1)}$ is achieved at $\alpha=\alpha_1$
through the evolution with $\upsilon=-\lambda\kappa$ and $\alpha_0=\alpha_2$~(\ref{eq:twospeeds})
started at $p_i^{(0)}=p_i^{(2)}$. With~(\ref{eq:p_approx}) one gets:
\begin{equation}
p_i^{(1)}=\tilde{p}_i(\alpha_1)+\lambda\kappa\tau_{R}^{}
\textstyle{\left(\frac{\mathrm{d}\tilde{p}_i}{\mathrm{d}\alpha}\right)_{\alpha_1}}+
o\left[\kappa\tau_{R}^{}\right];
\end{equation}
Treating in the same way $p_i^{(2)}$, one comes up with:
\begin{equation}
p_i^{(2)}=\tilde{p}_i(\alpha_2)-\kappa\tau_{R}^{}
\textstyle{\left(\frac{\mathrm{d}\tilde{p}_i}{\mathrm{d}\alpha}\right)_{\alpha_2}}+
o\left[\kappa\tau_{R}^{}\right];
\end{equation}

In the limit of very fast driving, $(\Delta\alpha/\kappa\tau_{R}^{}) \ll 1$, one has to
expand the exponents contained on the RHS of~(\ref{eq:turningpoints}) into power series.
Collecting terms up to first order in $\frac{\Delta\alpha}{\kappa\tau_{R}^{}}$ yields
the results for $p^{(1;2)}$ presented in Eq.~(\ref{eq:Corr_fast}). Their difference,
$\left(p^{(2)}-p^{(1)}\right)$, however, turns out to be of the second order and needs the
adequate accuracy in the treatment.

\section{\label{sec:no_change_over} Proof for $Q^*_{12}<0$}

In the stationary cyclic regime the system state evolution on the low-temperature stage $1\to 2$
is given by Eq.~(\ref{eq:walpha}) taken with $\upsilon=\kappa$, $\alpha_0=\alpha_1$ and $p_i^{(0)}=p_i^{(1)}$.
Let us denote this particular solution of the relaxation equation~(\ref{eq:alphder}) as $p_i^{\star}$; its
boundary values prove to be $p_i^{\star}(\alpha_{1;2})=p_i^{(1;2)}$ from~(\ref{eq:turningpoints}).
Consider the derivative of the sum $\sum\epsilon_i p_i^{\star}$ as a function $\varphi(\alpha)$:
\begin{equation}
\nonumber
\varphi(\alpha):=\frac{\mathrm{d}}{\mathrm{d}\alpha}\sum\epsilon_i p_i^{\star}=
-(\kappa\tau_R^{})^{-1}\sum \epsilon_i\left(p_i^{\star}-\tilde{p}_i(\alpha)\right),
\end{equation}
applying~(\ref{eq:alphder}).

It follows from~(\ref{eq:initial_phi}) that the boundary values of $\varphi(\alpha)$ turn out to be
$\varphi(\alpha_{1;2})\gtrless 0$, so there must be at least one $\alpha^*\in (\alpha_1;\alpha_2)$ such
that $\varphi(\alpha^*)=0$ and, consequently,
\begin{equation}
\nonumber
\sum\epsilon_i p_i^{\star}(\alpha^*)=\sum\epsilon_i \tilde{p}_i(\alpha^*).
\end{equation}
Choosing $\alpha^*$ as a new initial point for~(\ref{eq:walpha}), one can employ this equality
in order to recast $\sum \epsilon_i p_i^{\star}$ and then $\varphi(\alpha)$ into the new form:
\begin{eqnarray}
\label{eq:wstar}
\nonumber
\sum \epsilon_i p_i^{\star}&=&\sum \epsilon_i \tilde{p}_i(\alpha^*) e^{-\frac{\alpha-\alpha^*}{\kappa\tau_{R}}}\\
\nonumber
&+&(\kappa\tau_{R}^{})^{-1}\int_{\alpha^*}^{\alpha}\!
\left(\sum \epsilon_i \tilde{p}_i(\alpha^{\prime})\right)
e^{-\frac{\alpha-\alpha^{\prime}}{\kappa\tau_{R}}}\mathrm{d}\alpha^{\prime}\\
\nonumber
&=&\sum \epsilon_i \tilde{p}_i(\alpha)-
\int_{\alpha^*}^{\alpha}\!
e^{-\frac{\alpha-\alpha^{\prime}}{\kappa\tau_{R}}}\mathrm{d}\left(\sum \epsilon_i \tilde{p}_i\right);\\
\nonumber
\varphi(\alpha)&=&(\kappa\tau_R^{})^{-1}
\int_{\alpha^*}^{\alpha}\!
e^{-\frac{\alpha-\alpha^{\prime}}{\kappa\tau_{R}}}\mathrm{d}\left(\sum \epsilon_i \tilde{p}_i\right).
\end{eqnarray}
With~(\ref{eq:neg_deriv}) it is clear now that
\begin{equation}
\label{eq:phi_signs}
\varphi(\alpha)\gtrless0\quad\textrm{for}\quad\alpha\lessgtr\alpha^*,
\end{equation}
so $\alpha^*$ introduced above is unique.

Consider now the integral originated from~(\ref{eq:non-eq-heat})
\begin{equation}
\int^{\alpha_2}_{\alpha_1}\!\!
\alpha\,
\mathrm{d}\! \left(\sum\epsilon_i p_i^{\star}\right)\\
\nonumber
=
\int^{\alpha_2}_{\alpha_1}\!\!
\alpha\,
\varphi(\alpha)\,\mathrm{d}\alpha,
\end{equation}
which yields the total heat exchange $Q^*_{12}$ on the low-temperature isotherm
$g(\gamma)=\alpha\mathtt{T}_1$ of the Carnot cycle. It is convenient to split the
interval of integration by $\alpha^*$ so that the mean-value theorem becomes employable:
\begin{eqnarray}
\big(\!\int^{\alpha^*}_{\alpha_1}\!\!\!+\int^{\alpha_2}_{\alpha^*}\!\!\!\big)\,
\alpha\,
\varphi(\alpha)\,\mathrm{d}\alpha
&=&\theta_1\int^{\alpha^*}_{\alpha_1}\!\!\varphi(\alpha)\,\mathrm{d}\alpha
\nonumber\\
&+&\theta_2\int^{\alpha_2}_{\alpha^*}\!\!\varphi(\alpha)\,\mathrm{d}\alpha,
\nonumber
\end{eqnarray}
where $\alpha_1<\theta_1<\alpha^*<\theta_2<\alpha_2$. According to~(\ref{eq:phi_signs}), the
first addend here is positive while the second one is negative. One estimates further:
\begin{eqnarray}
\nonumber
\big(\theta_1\!\cdot\!\int^{\alpha^*}_{\alpha_1}\!\!\!&+&\,\,\theta_2\!\cdot\!\int^{\alpha_2}_{\alpha^*}\!\!\!\big)\,
\varphi(\alpha)\,\mathrm{d}\alpha\\
\nonumber
&<&\theta_2\int^{\alpha_2}_{\alpha_1}\!\!\varphi(\alpha)\,\mathrm{d}\alpha
  =\theta_2\int^{\alpha_2}_{\alpha_1}\!\!\mathrm{d}\left(\sum \epsilon_i p_i^{\star}\right)\\
\nonumber
&=&\theta_2\sum\epsilon_i\left(p_i^{(2)}-p_i^{(1)}\right)<0,
\end{eqnarray}
as it follows from~(\ref{eq:otto_heat}).

This completes the proof for
\begin{equation}
Q^*_{12}=
\int^{\alpha_2}_{\alpha_1}\!\!
\alpha\,
\mathrm{d}\! \left(\sum\epsilon_i p_i^{\star}\right)<0.
\end{equation}

\begin{acknowledgement}
One of us (J. B.) gratefully acknowledges financial and other support from DAAD and
the Russian Ministry for Education and Science in the framework of joint program
``Michail Lomonosov''. We thank Markus Henrich, Florian Rempp, and Georg Reuther
for valuable discussions.
\end{acknowledgement}


\begin{thebibliography}{29}


\bibitem{Bejan1988}
A.~Bejan, \emph{Advanced {E}ngineering {T}hermodynamics} (Wiley, N.Y., 1988)

\bibitem{TruesdellBharatha1977}
C.~Truesdell, S.~Bharatha, \emph{Classical {T}hermodynamics as a {T}heory of
  {H}eat {E}ngines} (Springer, New York, Berlin, 1977)

\bibitem{TodaKuboSaito1983}
M.~Toda, R.~Kubo, N.~Saito, \emph{Statistical {P}hysics {I}} (Springer Berlin,
  New York, 1983)

\bibitem{GemmerMichelMahler2004}
J.~Gemmer, M.~Michel, G.~Mahler, \emph{Quantum {T}hermodynamics} (Springer,
  2004)

\bibitem{Chamberlain2002}
R.V. Chamberlain, Science \textbf{298}, 1172 (2002)

\bibitem{Hill2001}
T.L. Hill, Nano Letters \textbf{1}, 111 (2001)


\bibitem{Alicki1979}
R.~Alicki, J. Phys. A \textbf{12}, L 103 (1979)

\bibitem{FeldmannKosloff2003}
T.~Feldmann, R.~Kosloff, Phys. Rev. E \textbf{68}, 016101 (2003)

\bibitem{GevaKosloff1992}
E.~Geva, R.~Kosloff, J.\ Chem.\ Phys. \textbf{97}, 4398 (1992)

\bibitem{JahnkeBirjukovMahler2007}
T.~Jahnke, J.~Birjukov, G.~Mahler, Eur. Phys. J. ST (conference proceedings) \textbf{151}, 167 (2007)

\bibitem{HenrichMahlerMichel2007}
M.~Henrich, G.~Mahler, M.~Michel, Phys.\ Rev.\ E \textbf{75}, 051118 (2007)

\bibitem{HenrichMichelMahler2006}
M.~Henrich, M.~Michel, G.~Mahler, Europhys. Lett. \textbf{76}, 1058 (2006)

\bibitem{PalaoKosloffGordon2006}
J.P. Palao, R.~Kosloff, J.~Gordon, Phys. Rev. E \textbf{64}, 056130 (2001)

\bibitem{QuanZhangSun2006}
H.T. Quan, P.~Zhang, C.P. Sun, Phys. Rev. E \textbf{73}, 036122 (2006)

\bibitem{ScovilSchulz1959}
H.E.D. Scovil, E.O. Schulz-Du~Bois, Phys. Rev. Lett. \textbf{2}, 262 (1959)

\bibitem{SegalNitzan2006}
D.~Segal, A.~Nitzan, Phys. Rev. E \textbf{73}, 026109 (2006)

\bibitem{TonnerMahler2005}
F.~Tonner, G.~Mahler, Phys.\ Rev.\ E \textbf{72}, 066118 (2005)

\bibitem{HackermuellerHornberger2004}
L.~Hackerm\"uller, K.~Hornberger, Nature \textbf{427}, 711 (2004)

\bibitem{CurAhlb1975}
F.L. Curzon, B.~Ahlborn, Am.\ J.\ Phys. \textbf{43}, 22 (1975)

\bibitem{Sheehan2002}
D.P. Sheehan, AIP Conf. Proc. (AIP Press, Melville, N.Y.) \textbf{643} (2002)

\bibitem{Jones}
R.~Jones, www.softmachines.org/wordpress/?p=127

\bibitem{Cheng2005}
T.P. Cheng, \emph{Relativity, {G}ravitation and {C}osmology} (Oxford U. P.,
  2005)

\bibitem{RemppMichelMahler2006}
F.~Rempp, M.~Michel, G.~Mahler, Phys.\ Rev.\ A \textbf{76}, 032325 (2007)

\bibitem{Kieu2004}
T.D. Kieu, Phys. Rev. Lett. \textbf{93}, 140403 (2004)

\bibitem{BenderBrodyMeister2002}
C.M. Bender, D.C. Brody, B.K. Meister, Proc. Royal Soc. (London) \textbf{A
  458}, 1519 (2002)


\bibitem{vdBroek2005}
C.V.d. Broeck, Phys.\ Rev.\ Lett. \textbf{95}(19), 190602 (2005)

\bibitem{IzumidaOkuda2008}
Y. Izumida, K. Okuda, arXiv:0802.3759 (2008)

\bibitem{Tsallis1988}
C.~Tsallis, J. Stat. Phys. \textbf{52}, 169 (1988)

\end{thebibliography}

\end{document}